\documentclass[a4paper,12pt]{article}
\usepackage[]{inputenc,graphicx,amsmath,textcomp,amssymb}
\usepackage{color}
\usepackage{subfigure}  
\usepackage{hyperref}
\usepackage[margin=1.15in,top=1.1in]{geometry}

\newcommand{\eqi}{\begin{equation}}
\newcommand{\eqf}{\end{equation}}
\newcommand{\bea}{\begin{eqnarray}}
\newcommand{\eea}{\end{eqnarray}}

\begin{document}
\begin{flushright}
 SINP-APC-12/3 
\end{flushright}

\begin{center}
{\Large{\bf 
Gamma Ray and Neutrino Flux from Annihilation of Neutralino Dark Matter at Galactic Halo Region in mAMSB Model 
}}\\
\vspace {1.0cm}
{Kamakshya Prasad Modak\footnote{email: kamakshya.modak@saha.ac.in } and 
Debasish Majumdar\footnote{email: debasish.majumdar@saha.ac.in}}\\
\vspace{0.40cm}
\it{ Astroparticle Physics and Cosmology Division, }\\
\it{Saha Institute of Nuclear Physics,}\\
\it{ 1/AF Bidhannagar, Kolkata 700064, India.}
\end{center}


\begin{abstract}
We consider the lightest supersymmetric particle (LSP), neutralino in 
minimal anomaly mediated supersymmetry breaking model 
(mAMSB) to be
a possible candidate for
weakly interacting massive particles (WIMP) or cold dark matter and investigate
its direct and indirect detections. 
The theoretically allowed supersymmetric parametric space for such a model 
along with the recent bounds from LHC is
constrained by the WMAP
results for relic densities.
The spin independent and 
spin dependent scattering cross sections for dark matter off nucleon
are thus constrained from the WMAP results. They are found to be within the 
allowed regions of different ongoing direct detection experiments.
The annihilation of such dark matter candidates at the galactic centre 
produce different
standard model particles such as gamma rays, neutrinos etc. 
In this work, we calculate the possible fluxes 
of these $\gamma$-rays and neutrinos coming
from the direction of the
galactic centre (and its neighbourhood) at terrestrial or satellite borne detectors.
The calcutated $\gamma$-ray flux is compared with the observational results
of HESS experiment.
The neutrino flux of different flavours from the galactic centre and at different
locations away from the galactic centre 
produced by WIMP annihilation in this model are also 
obtained for four types of galactic dark matter halo profiles.
The detection prospects of such $\nu_\mu$ coming from
the direction of the galactic centre at the ANTARES under sea detector
are discussed in terms of muon signal yield from these muon neutrinos.
Both the gamma and neutrino signals are estimated for four different
dark matter halo profiles.

\end{abstract}

\newpage

\section{Introduction}

Cosmological observations like flattening of 
rotation curves of spiral galaxies \cite{rotation}, the gravitational 
microlensing \cite{gravitationallensing}, observations on Virgo and Coma clusters \cite{virgo,coma}, 
bullet clusters \cite{bullet},
etc. provide indications
of existence of huge amount of non-luminous matter or 
dark matter (DM) in the universe. The
Wilkinson Microwave Anisotropy Probe (WMAP) experiment \cite{wmap} 
suggests that more then 80\% of the
total matter content of the universe 
(almost 23\% of the total content of the universe) is dark matter.
The general wisdom is that in order to account for
the relic abundance of DM, a candidate for dark matter
should be massive, very weakly interacting
and non-relativistic (cold dark matter or CDM) particles.
This allows the structure formation on large scales.
In the present work, we consider such weakly interacting massive
particles (WIMPs) \cite{jungman,griest,bertone,murayama} to
consist of the total DM content
of the universe.

Because of its nature, the detection of dark matter is very challenging
experimental effort. In general there are two types of detection 
mechanism namely direct detection of dark matter and indirect
detection of dark matter.
The indirect detection of dark matter involves detecting the
particles (and their subsequent decays) or photons produced due to  
dark matter annihilations. These annihilation products can be 
fermions or $\gamma$ photons. The dark matter particles, if 
trapped by the gravity of a massive body like sun or galactic 
centre, can annihilate there to produce these particles. Study of
such photons and fermions such as neutrinos thus throw light on the
nature of galactic dark matter as well as the nature of the galactic
dark matter
halo profile. 
Different satellite-borne and ground-based experiments looking for 
extra terrestrial gamma signals have reported the observence of excess gamma ray signals
in the direction of galactic centre in different energy regions.
If the observed TeV gamma rays from the galactic centre are indeed due to
the annihilation of dark matter at galactic centre then such dark matter mass should be
$\sim$ TeV. The HESS (The High Energy Stereoscopic System) \cite{hess1,hess2} experiment had reported 
the gamma rays from the galactic centre with energies in TeV range.
In minimal anomaly mediated supersymmetry breaking (mAMSB) model \cite{randall,guidice},
the lightest supersymmetric particle (LSP) neutralino that can be 
a candidate for dark matter has its mass in TeV range.
The calculated $\gamma$-ray flux is found to be within the experimental
search limit of high energy gamma ray search experiments such as HESS.


The dark matter candidate in the present work is considered to be
the lightest supersymmetric particle (LSP) neutralino in the 
minimal anomaly mediated supersymmetry 
breaking model where the LSP is stabilised by conservation of R-parity.
In the superconformal Anomaly Mediated Supersymmetry Breaking (AMSB) 
mechanism, dynamical or spontaneous breaking is supposed to
take place in some `hidden' sector (HS) and this breaking
is mediated to the observable sector (OS) by gravitino mass 
($m_{\frac{3}{2}}$) $\sim 100$ TeV. 
Supersymmetry breaking effects in the observable sector
have a gravitational origin in this framework.
In ordinary gravity-mediated supersymmetry breaking model,
the supersymmetry breaking is transmitted from
HS to OS via tree level exchanges with gravitational coupling.
But in AMSB, the HS and the OS superfields are assumed to be
located in two parallel but distinct 3-branes and the 3-branes
are separated by bulk distance which is of the
order of compactification radius, $r_c$.
Thus any tree level exchange with mass higher than the inverse of
$r_c$ is exponentially suppressed. So, the supersymmetry breaking
is propagated from the HS to the OS via
loop generated superconformal anomaly.

In AMSB model, the slepton mass-squared terms are 
negative giving to
tachyonic states. The problem is circumvented by adding 
an universal mass-squared term $m^2_0$ to all the
squared scalar masses in the minimal extension to this theory, namely,
minimal anomaly mediated supersymmetry breaking (mAMSB)
model \cite{randall,guidice}.
An sparticle spectrum in this model is fixed by three parameters, 
$m_{\frac{3}{2}}$ which is gravitino mass, tan$\beta$ which
is the ratio of the vacuum expectation values of the two Higgs 
fields ($H^0_1$ and $H^0_2$) and sign($\mu$), where $\mu$ 
is the Higgsino mass. Thus four parameters are needed
to generate spectrum in mAMSB.
The neutralino is the lowest mass eigenstate of linear superposition
of photino ($\tilde{\gamma}$), zino ($\tilde{Z}$),
and the two Higgsino states ($\tilde{H^0_1}$ and $\tilde{H^0_2}$) 
\cite{haber}, written as, 
\begin{equation}
\chi = a_1\tilde{\gamma} + a_2\tilde{Z} + 
a_3\tilde{H^0_1} + a_4\tilde{H^0_2}\,\,\, .
\end{equation}
in the basis $ \left(\begin{matrix}
 \tilde{\gamma} & \tilde{Z} &\tilde{H^0_1}  & \tilde{H^0_2}
             \end{matrix}\right) . $          

The ATLAS collaboration \cite{atlas} has recently performed an improved
analysis and give a new constraint on the
chargino mass to $\sim$ 118 GeV. This new constraint differs from the previous
LEP2 bound.
In this work, the SUSY parameter space namely $m_0$, $m_{\frac{3}{2}}$,
tan$\beta$ and sign($\mu$) is initially adopted from Datta {\it et al.}
\cite{datta} but with proper incorporation of the recent LHC 
(ATLAS) bound on chargino mass \cite{atlas} mentioned above.
The relic  
densities for such dark matter are then computed using 
these SUSY parameters and they are compared with the WMAP results.
The parameters, thus constrained further by the WMAP results, are then used to calculate the 
spin independent and spin dependent cross sections ($\sigma_{\rm scatt}$) 
for different neutralino masses ($m_{\chi}$) (obtained using the 
restricted parameter space). The 
$\chi$-nucleon scattering process is  
essential for the direct searches of dark matter. 
As mentioned above, we calculate $\chi$-nucleon elastic scattering
cross section $\sigma_{\rm scatt}$ for the restricted parameter space
discussed earlier.
The $m_{\chi}-\sigma_{\rm scatt}$ region, thus obtained, 
is found to be within the allowed limits of most of the direct detection 
experiment results.

Using the constrained 
mAMSB parameter space discussed above we calculate the gamma ray flux 
in the direction of the galactic centre. These studies are performed 
for different galactic 
dark matter halo profiles.     
We find that the gamma spectrum from galactic centre and halo
produced by neutralino dark matter 
within the framework of the present mAMSB model,  
is highly energetic. The experiment like  
HESS, 
that can probe high energy gamma rays and
which, being in the southern hemisphere has
better visibility of the galactic centre,
will be suitable to test
the viability of the present dark matter candidate in mAMSB model.
The possibility of detecting neutrinos from galactic 
centre and halo from 
dark matter annihilations are also addressed with reference 
to ANTARES (Astronomy with a Neutrino Telescope and Abyss environmental RESearch) 
\cite{antares} under sea neutrino experiment.   

In a recent work V\'{a}squez {\it et al.} \cite{prd84nmssm} has given a detailed
analysis of the allowed parameter space for a neutralino dark matter 
in the framework of NMSSM model. In their case the dark matter (neutralino)
mass was within the range of $\sim$ 80 GeV and hence the energies of the 
gamma rays from 
such dark matter annihilations can be probed by FermiLAT \cite{fermilat}
experiment. In the present calculation, we instead consider the 
neutralino dark matter in mAMSB model mentioned above. 
Some of the earlier works on dark matter phenomenology in AMSB 
model include Baer {\it et al.} \cite{baer}, Moroi {\it et al.} \cite{moroi},
Ullio \cite{ullio} etc. In Refs. \cite{baer} and \cite{ullio}
the $\gamma$ flux from the galactic centre are discussed 
and although neutrinos from the neutralino annihilations are mentioned 
in Ref. \cite{baer} but they have not discussed elaborately. 
Moreover 
only two halo models are considered for their analysis. In an another 
earlier work (\cite{dm}), a neutralino dark matter in AMSB model 
is studied to obtain the region in   
scalar cross section ($\sigma_{\rm scatt}$ - $m_\chi$) parameter 
space. But in this case  WMAP limit has not been taken into account.     
In Ref. \cite{hooper}, the $\gamma$ signal from galactic centre region 
due to dark matter annihilation is addressed mainly 
for the case of FERMI (formerly GLAST \cite{glast}) satellite-borne experiment.
Ref. \cite{serpico} discusses the the $\gamma$-flux from 
galactic centre region, originated by dark matter annihilations. 
The authors made the analysis with different particle dark matter 
candidates with reference to MSSM, Kaluza-Klein extra dimensional model 
etc. for different halo profiles and taking into account the Fermi-LAT
experiment. But the neutrinos as dark matter annihilation products 
are not addressed. In another work by Allahverdi et al \cite{bhaskar}
considered MSSM and $U(1)_{B-L}$ extened MSSM 
model for dark matter candidate and calculated 
$\gamma$ and neutrino fluxes
from galactic and extra-galactic origins by annihilating dark matter.
But they have considered only one dark matter halo profile namely NFW halo profile and 
they have not shown the neutrinos flux for different neutrino 
flavours. Moreover, no detailed comparison of their results with 
high energy neutrino or gamma search experiments is shown.
There are also other earlier works like \cite{klophov} where dark matter
annihilations in galaxy are addressed.   
  
In this work we use the mAMSB framework for the neutralino DM
candidate and study both the possible $\gamma$-ray and neutrino flux
that an experiment will probe in the direction of galactic centre.
We perform this study for four dark matter halo profiles.
The $\gamma$-ray results are compared with HESS experiment and for 
neutrinos, we estimate the possible signal in ANTARES under sea detector.

The paper is organised as follows. In section 2 we discuss the 
calculation of relic densities of mAMSB neutralinos for the 
parameter space. The relic 
densities are then compared with the WMAP results. The
parameter space thus constrained further by WMAP is then used 
to calculate the spin dependent and spin independent scattering 
cross sections. They are compared with the existing direct detection 
experiment limits. These are discussed in section 3. In section 
4 the indirect detection of the mAMSB dark matter from their 
annihilations at galactic centre and halo are discussed. To this end 
the gamma signals and neutrino signals are addressed. Finally 
in section 5 we give discussions and conclusions.

\section{Relic Abundance Calculation}

In order to calculate the relic abundance of the LSP, $\chi$, 
one needs to consider annihilation of $N$ supersymmetric
particles with masses $m_i$ ($i$=1,2,..,$N$) and internal degrees
of freedom $g_i$ respectively.
The relic abundance is obtained by numerically solving the Boltzmann's 
equation,
\bea
\frac{dn}{dt} + 3Hn = -\langle\sigma v\rangle (n^2-n_{\rm eq}^2)\,\, ,
\label{5}
\eea
where $n$ is the total number density of all the supersymmetric particles $n_i$ 
$$n=\Sigma_i n_i\,\,\, ,$$
and 
$n_{\rm eq}$ is the value of $n$ when the particles for dark matter 
candidate were in chemical equilibrium. 
At this epoch the temperature $T$ of the universe was greater than $T_f$ 
($T > T_f$),
the freeze out temperature of the particle considered. 
At a temperature below the freeze-out temperature $T_f$, the particles 
falls out of chemical 
and thermal equilibrium and their co-moving number density becomes 
fixed or ``frozen". 
In Eq. \ref{5}, $H$ denotes the Hubble parameter and  
$\langle\sigma v\rangle$ is the thermal average of the product of
annihilation cross section and the relative velocity of the two
annihilating particles. 
\begin{eqnarray}
 \langle\sigma v\rangle &=& \sum_{i,j}\langle\sigma_{ij} v_{ij}\rangle\frac{n^{(i)}_{\rm eq} n^{(j)}_{\rm eq}}{n^2_{\rm eq}}\,\,\, ,
\end{eqnarray}
with $$v_{ij} = \frac{\sqrt{(p_i.p_j)^2-m^2_im^2_j}}{E_iE_j}\,\, .$$
In the above, ($p_i,p_j$) and ($E_i,E_j$) are the momenta and energies 
respectively for the $i$th and $j$th particles.  
Defining the abundance,
$Y = n/s$ \cite{gondolo} where $s$ is the total entropy density 
of the universe, and with the dimensionless quantity $x = m_{\chi}/T$, with
$m_{\chi}$ being the mass of LSP, 
Eq. \ref{5} can be written 
in the form 
\begin{equation}
\frac{dY}{dx} = \frac{1}{3H}\frac{ds}{dx}\langle\sigma 
v\rangle(Y^2-Y_{eq}^2)\,.
\label{6}
\end{equation}
In Eq. \ref{6}, $Y_{\rm eq}$ is the value of $Y$ when $n = n_{\rm eq}$. 
With Hubble parameter $H = \sqrt {\frac{8}{3}\pi G\rho}$,  
$G$ being the gravitational constant, the total energy density ($\rho$) 
and the total entropy density ($s$) of the universe are given by
\cite{gondolo}
\bea
\rho &=& g_{eff}(T)\frac{\pi^2}{30}T^4 
\label{a}
\eea
\bea
{\rm and}\,\,\,\,\, s &=& h_{eff}(T)\frac{2\pi^2}{45}T^3\,\,\, .  
\label{b}
\eea
In Eqs. \ref{a} and \ref{b} $g_{eff}$, $h_{eff}$ are the effective 
degrees of freedom for the energy and entropy densities respectively. 
Substituting Eqs. \ref{a}, Eqs. \ref{b} and the expression for $H$ 
in Eq. \ref{6}, one obtains the evolution equation  
of $Y$ as
\begin{equation}
\frac{dY}{dx} = -\left( \frac{45}{\pi}G \right )^{-1/2}\frac{g_*^{1/2}m_{\chi}}{x^2}
\langle \sigma v \rangle (Y^2-Y_{eq}^2)\,\,\, ,
\label{8}
\end{equation}
where $g_*^{1/2}$ is defined as \cite{gondolo}
\begin{equation}
g_*^{1/2} = \frac{h_{eff}}{g_{eff}^{1/2}}
\left (1+\frac{1}{3} \frac{T}{h_{eff}}\frac{dh_{eff}}{dT} \right)\, .
\end{equation}
The expression for $Y_{eq}$ is given by \cite{gondolo}
\begin{equation}
Y_{eq}(T) = \frac{45}{4\pi^4h_{eff}(T)}\sum_i{g_i\frac{m_i^2}{T^2}}K_2\left(\frac{m_i}{T}\right)\,\,\, ,
\label{9}
\end{equation}
where we sum over all supersymmetric particles denoted by $i$ 
with mass 
$m_i$ and internal degrees of freedom $g_i$.
$K_2(x)$ is the modified bessel function 
of the second kind of order $2$. 
The thermally-averaged cross
section, $\langle\sigma v\rangle$ must include all channels by which $\chi$ 
can interact,
including coannihilation with other particles, in
which the number densities of both species are important. 

Integrating Eq. \ref{8} from $x = x_0 = m/T_0$ to $x = x_f = m/T_f$, 
where $T_0$ is the present photon temperature (2.726$^o$ K)
we obtain $Y_0$ (value of $Y$ at $T = T_0$) which is needed to 
compute the relic density.
Eq. \ref{8} is solved numerically with the following 
approximations, 
\begin{itemize}
\item{1.} At small $x$ (high $T$), the abundance of lightest 
SUSY particles (LSP) 
are almost in equilibrium and the temperature
variation of the deviation from equilibrium abundance is negligible, 
i.e., $Y\approx Y_{eq}$ and $\frac{Y-Y_{eq}}{T}\approx 0$.
Thus, the evolution equation reduces to,
\begin{equation}
 \frac{dln(Y_{eq})}{dx}= -\left( \frac{45}{\pi}G \right )^{-1/2}\frac{g_*^{1/2}m_{\chi}}{x^2}
\langle \sigma v \rangle Y_{eq}\delta(\delta+2)\,\,\, ,
\end{equation}
where $\delta$ is some small constant coming from the definition of 
freeze-out temperature $T_f$.
\item{2.} At temperature below $T_f$, equilibrium abundance, 
$Y_{eq}$ falls much 
below $Y$, as seen from Eq. \ref{8} and can be neglected in
the abundance evolution equation. Thus, $Y_0$ is obtained 
from the relation,
\begin{equation}
 \frac{1}{Y_0}=\frac{1}{Y_f}-m_{\chi}\left( \frac{45}{\pi}G \right )^{-1/2}\int^{x_0}_{x_f}\frac{g_*^{1/2}(x)}{x^2}
\langle \sigma v \rangle dx
\end{equation}
\end{itemize}
The relic density of LSP, in the units of critical density, 
$\rho_{cr}=3H^2/8\pi G$, can be expressed as
\begin{equation}
 \Omega_{\chi}=\frac{m_{\chi}n}{\rho_{cr}}=\frac{m_{\chi}s_0Y_0}{\rho_{cr}}\,\,\, ,
\end{equation}
where $s_0$ is the present entropy density evaluated at $T_0$. 
Finally, knowing $Y_0$, we can compute the relic density of the
dark matter candidate,
from the relation \cite{gondolo},
\bea
\Omega_\chi h^2 &=& 2.755\times10^8 \frac{m_{\chi}}{\rm GeV}Y_0\,\,.
\label{}
\eea
In the above
$h$ is the Hubble constant in $100\,{\rm Km}\,\,{\rm sec}^{-1}{\rm Mpc}^{-1}$ 
unit.
The WMAP survey combining with recent observations of large–scale 
structure provides the  
constraints on the dark matter density $\Omega_{DM} h^2$ as  
\begin{equation}
 0.099 < \Omega_{DM}h^2 < 0.123 \,\,\, .
\end{equation}
where $\Omega_{DM}$ is
the ratio of dark matter density to the critical density 
$\rho_c = 1.88h^2\times10^{-29}{\rm g}{\rm cm}^{-3}$. 

In the present work we calculate the relic densities for the dark 
matter candidate neutralino in mAMSB model and compare our results 
with the WMAP bound. The allowed parameter space in the 
present SUSY model is thus extracted by WMAP results.  

\begin{figure}[h]
\begin{center}
\includegraphics[width=2.7in,height=4.5in,angle=-90]{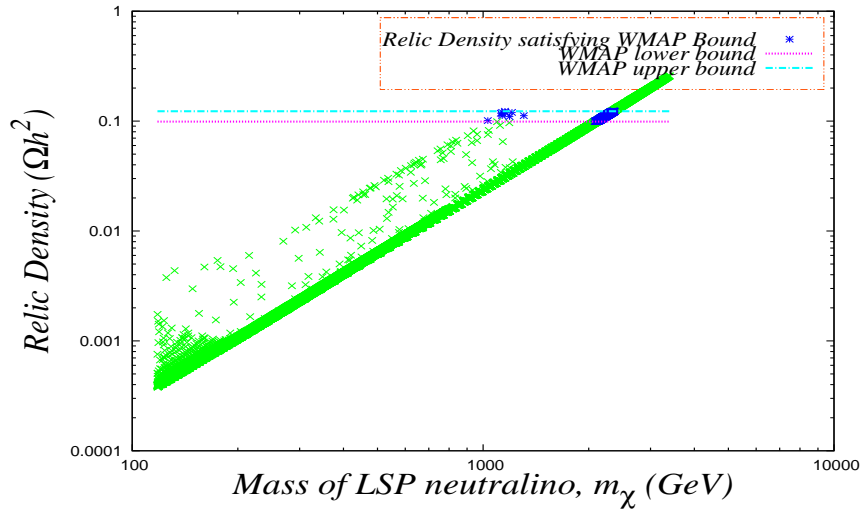}
\caption{\label{fig:relic} \textit{Scatter
plot of mass of the LSP neutralino ($m_{\chi}$) vs. 
relic density ($\Omega h^2$) in mAMSB model. The cyan and pink line
represents the WMAP upper and lower bounds on dark matter
relic density respectively and the blue dotted zones
corresponds to the mass range satisfying the WMAP
limits.}}
\end{center}
\end{figure}

As mentioned earlier, the parameter space of the  
mAMSB model is defined by the four parameters, namely
$m_{3/2}$, $m_0$, $\tan\beta$ and sign$(\mu)$.
The whole parameter space defined by the above parameters 
and constrained
by the allowed region of $m_0-m_{3/2}$ (see earlier) 
is used to calculate the relic density $\Omega_\chi$ (or $\Omega_\chi h^2$)
and the results are then compared with the WMAP results.

The relic density in the present formalism of SUSY model is
computed using the code 
\texttt{micrOMEGAs} \cite{micromegas}. We thus obtain 
the relic density for the scanned SUSY parameter space
discussed above.
We find that the generated LSP neutralinos span very large range of
mass.
Each generated LSP neutralino mass gives rise to different annihilation 
cross section due to their annihilations to
different standard model particles and also co-annihilation processes. 
We mention here that the LSP neutralino is found to be wino dominated
with the other components like bino or higgsino have very negligible contribution.  
The mass scales for other sparticles are above the LSP neutralino
mass scale. For example for an LSP of mass $\sim 2$ TeV, the sneutrinos mass is $\sim$ 14 TeV and
for squark the mass scale is $\sim$ 18 TeV; the NLSP mass is $\sim$ 7 TeV.

In Fig. \ref{fig:relic}, the variation of relic densities for different LSP 
neutralino masses are shown.
The scatter plots in Fig. \ref{fig:relic} correspond to
the allowed parameter space. The WMAP bound is superimposed on this scatter
plot in Fig. \ref{fig:relic} and the regions of agreement of the present
calculational results with WMAP data are identified by blue coloured
area in Fig. \ref{fig:relic}.
From Fig. \ref{fig:relic}, we obtain two 
different neutralino mass regions satisfying the WMAP bound. 
One region is around 1 TeV and
the other region is at a somewhat higher range of $\sim$ 2 TeV.

\begin{figure}[h]
\begin{center}
\includegraphics[width=2.7in,height=4.5in,angle=-90]{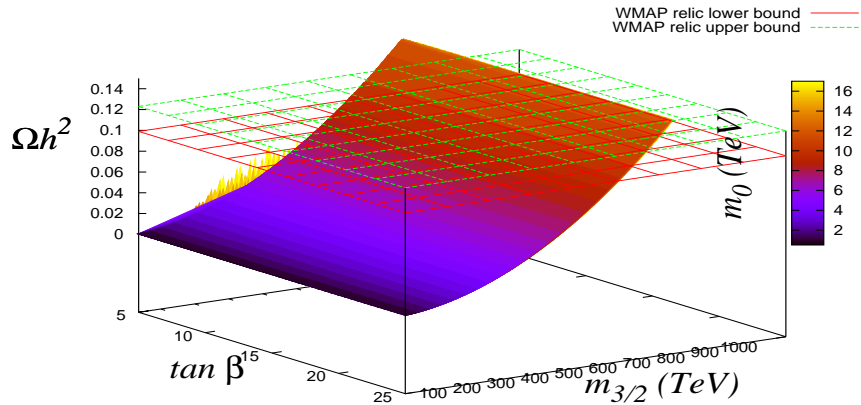}
\caption{\label{paramscan} \textit{Constraints on SUSY 
parameter space from WMAP limits in present SUSY model. 
The gaugino mass parameter $m_0$ are shown by the colour index
where $m_0$ varies from blue coloured region to yellow 
region as its mass increases.}}
\end{center}
\end{figure}

In order to elaborate how the WMAP bound 
constrains the SUSY parameter space in the present mAMSB model, we make 
a 3-D colour coded plot in Fig. \ref{paramscan}, where the variation 
of relic density
$\Omega h^2$ with the simultaneous variations of all three SUSY 
parameters namely $m_{3/2}$, $\tan \beta$ and $m_0$ are furnished. 
In Fig. \ref{paramscan}, the parameters $m_{3/2}$ and $\tan \beta$ are plotted 
along X and Y axes respectively while the variation of gaugino mass 
$m_0$ is shown in colour coded display whereby the colour reference
deep blue denotes the lower value of $m_0$ and increases towards 
the yellow zone in the plot. The corresponding variation of 
$\Omega h^2$ is shown along Z axis. The WMAP limits 
are shown in Fig. \ref{paramscan} 
by two meshes separated by the WMAP limit along $\Omega h^2$ axis. 
One observes from Fig. 2 that a very small region of the 
$m_{3/2}-m_0-\tan \beta$ parameter space is allowed by 
WMAP. Thus WMAP limit further constraints the $m_{3/2}-m_0$ parameter
limits.  
From Fig. \ref{paramscan} it is also clear that only higher values 
of $m_0$ ($\sim 10 - 12$
TeV), and $m_{3/2}$ ($\sim 650 - 700$) TeV could satisfy the WMAP limits.
We have not obtained any other parameter space in $m_0 - m_{3/2}$ plane  
that satisfy WMAP limits.

\begin{figure}[h]
\begin{center}
\includegraphics[width=2.7in,height=4.5in,angle=-90]{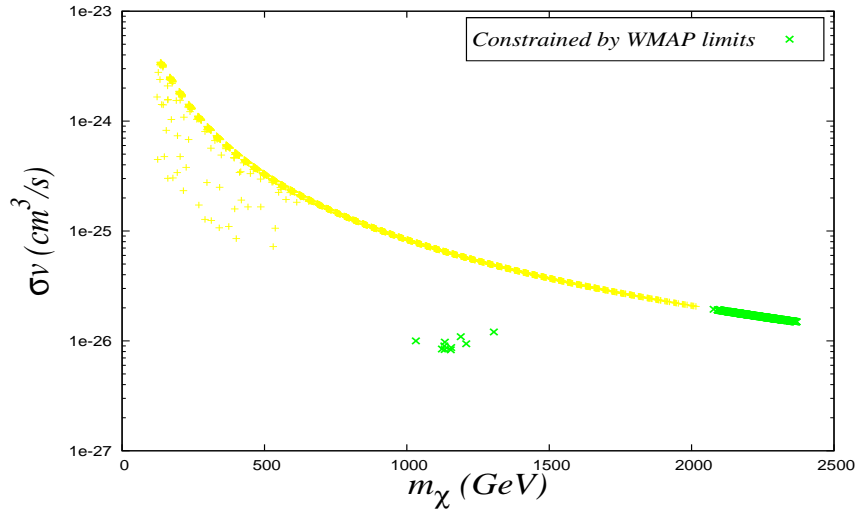}
\caption{\label{sigma} \textit{Plot showing the variation (in yellow) of
annihilation cross section times relative velocity of annihilating
neutralinos ($\sigma v$) with the mass of the LSP neutralino
($m_{\chi}$) in mAMSB model. The
green zones are the WMAP allowed regions.}}
\end{center}
\end{figure}

In Fig. \ref{sigma}, we show how the annihilation cross sections
vary with the neutralino dark matter mass ($m_\chi$) in the 
present model. The WMAP allowed mass region is also shown by green 
colour. The $\sigma v$ for the allowed zones
(marked green) are seen to be around the value 
$\sim 10^{-26}$ cm$^{3}$sec$^{-1}$.

\begin{figure}[h]
\begin{center}
\includegraphics[width=2.6in,height=4.5in,angle=-90]{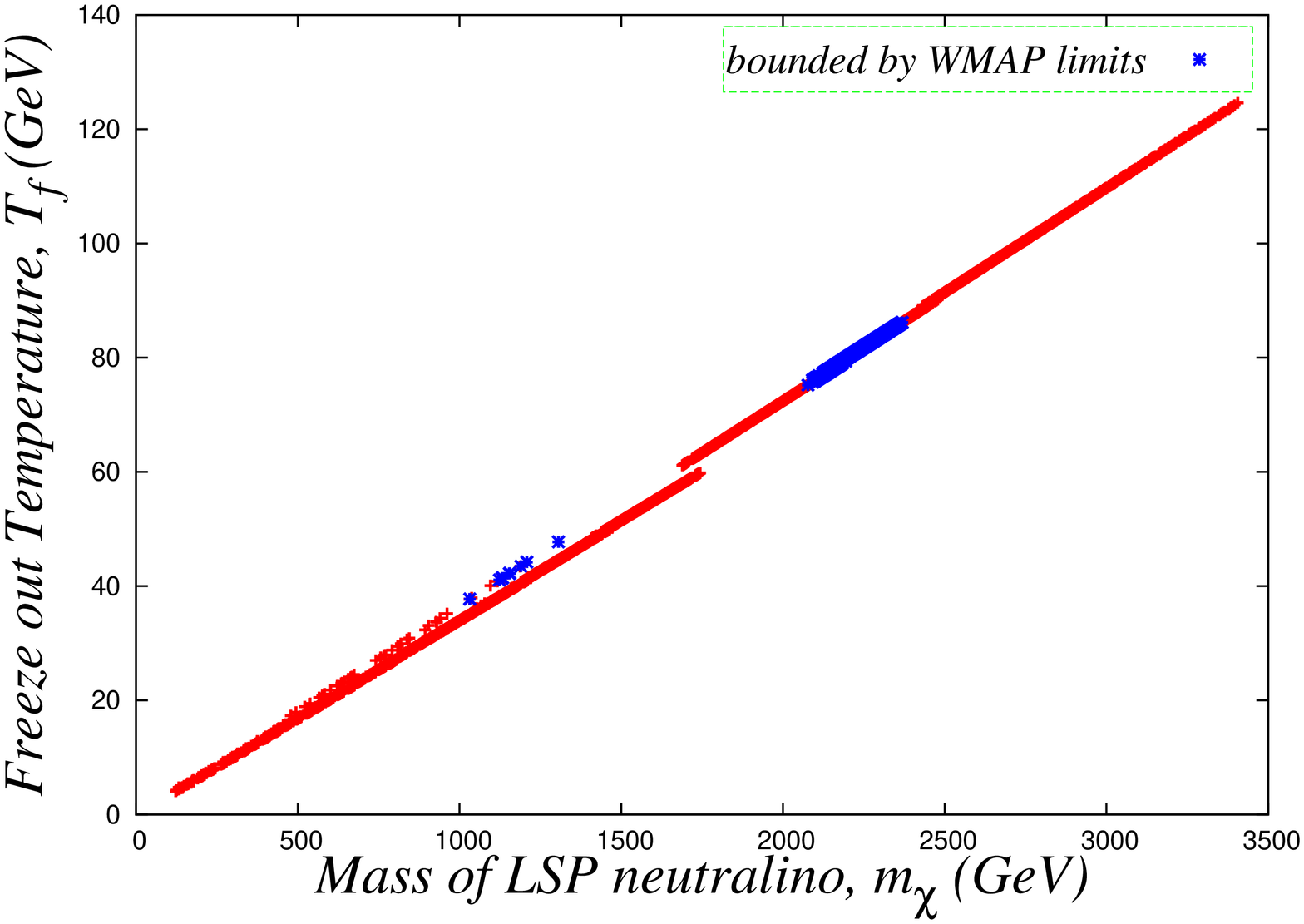}
\caption{\label{fig:freezeout} \textit{Plot showing 
the variation (in red) of freezeout temperature ($T_f$)
with the mass of the mAMSB LSP neutralino 
($m_{\chi}$). The blue dotted zones are constrained by 
WMAP limits on dark matter relic.}}
\end{center}
\end{figure}

The variations of freezeout temperatures ($T_f$) of LSP neutralino 
for the mass range obtained 
using the SUSY parameter space discussed earlier, are shown in  
the scatter plot of Fig. \ref{fig:freezeout}. 
The neutralinos that satisfy WMAP relic density results are shown  
as blue in Fig. \ref{fig:freezeout}. As in Fig. 1, in this case also 
one observes two such regions, one is around $T_f \sim 80-86$ GeV 
(more populated) and 
the other (fewer candidates) is at a lower region of 
around $T_f \sim 40$ GeV.

\section{Direct Detection}

The direct detection of dark matter is based on the principle that
the WIMP scatters off the target nucleus of the material of the detector  
causing the nucleus to recoil.
The signal generated
by the nuclear recoil (generally $\sim$ keV) is measured for direct 
detection.
In the direct detection experiments, attempts are made to 
give a bound in the 
$m_\chi - \sigma_{\rm scatt}$ space ($m_\chi$ being the mass of the dark matter 
and $\sigma_{\rm scatt}$ is the dark matter-nucleus or dark matter-nucleon
scattering cross sections). 
Different techniques are adopted by different direct detection experiments
in order to measure the nuclear recoil energies. Some experiments
that use Ge, Si or NaI as detector materials use scintillation,
phonon or ionization techniques. In another class of detectors like 
Time Projection Chamber or TPC detectors, the drifting of 
ionized charges, produced by recoil nucleon of the detector material (generally 
noble liquids like xenon, argon and neon), produce the track from 
which the direction of recoil can also be measured.   
Some of the ongoing direct detection experiments include
DAMA (NaI) \cite{dama}, CDMS ($^{73}$Ge) \cite{cdms,cdms2} ,
PICASSO (CS$_2$) \cite{picasso} , XENON \cite{xenon10,xenon100}, 
COUPP \cite{coupp}, LUX  (use xenon) \cite{lux}, CLEAN (use liquid argon and neon
as scintilator) and DEAP (use argon) \cite{deapclean} etc. They give 
different limits on scattering cross sections for different dark matter mass. 

The dark matter-nucleus
scattering cross sections  
can be of two types namely axial-vector (spin-dependent) or scalar 
(spin-independent). The target nucleus, with zero ground state spin 
gives rise to spin independent interaction.
On the other hand, spin-dependent
interactions are for the nuclei with unpaired nucleon 
that gives rise to non-zero ground state spin.
The experiments such as Edelweiss \cite{edelweiss}, DAMA/NaI , CDMS 
SuperCDMS \cite{supercdms}, Xenon10 \cite{xenon10},
Xenon100 \cite{xenon100}, Zeplin \cite{zeplin,zeplin3}, KIMS \cite{kims}, CoGeNT \cite{cogent}
are using detectors made of heavy nuclei (Ge or Xe) to search 
scalar interactions. On the other hand, NAIAD \cite{naiad}, 
SIMPLE \cite{simple}, PICASSO,
Tokyo/NaF \cite{tokyonaf}
are using light nuclei to detect spin-dependent case.

The interaction Lagrangian for spin independent elastic scattering
of Majorana fermionic WIMP off nucleon $N$ in non-relativistic
limit is given by \cite{falk},
\begin{eqnarray}
 L_{SI} &=& \lambda_N\bar\psi_\chi\psi_\chi\bar\psi_N\psi_N\,\,\, ,
\end{eqnarray}
where $\lambda_N$
is the WIMP-nucleon coupling. Other notations have their usual significance.
The interaction Lagrangian for spin-dependent case 
is given by \cite{falk},  
\begin{eqnarray}
L_{SD}&=&
\epsilon_N\bar\psi_\chi\gamma_\mu\gamma_5\psi_\chi\bar\psi_N
\gamma^\mu\gamma_5\psi_N\,\, , \label{Lsd}
\end{eqnarray}
where $\epsilon_N$ denotes the coupling.
The spin-dependent and spin-independent cross sections for scattering of dark matter 
particle ($\chi$) with nucleon ($N$) are
respectively given in compact forms as,
\begin{equation}
\sigma^{\rm SD}=\frac{4m_{\chi}^2 M_N^2}{\pi(m_{\chi}+M_N)^2}\times3|A^{\rm SD}|^2 \,\,\, ,
\end{equation}
\begin{equation}
\sigma^{\rm SI}=\frac{4m_{\chi}^2 M_N^2}{\pi(m_{\chi}+M_N)^2}\times|A^{\rm SI}|^2 \,\,\, ,
\end{equation}
where $m_{\chi}$, $M_N$ are the dark matter particle mass and nucleon 
mass respectively. In the above, $A^{\rm SI}$ and $A^{\rm SD}$ are the relevant matrix elements
that depend on the quark contents of the target nucleon ($N$)
for $\chi$-$N$ scattering.

\begin{figure}[h]
\begin{center}
\includegraphics[width=2.5in,height=4.2in,angle=-90]{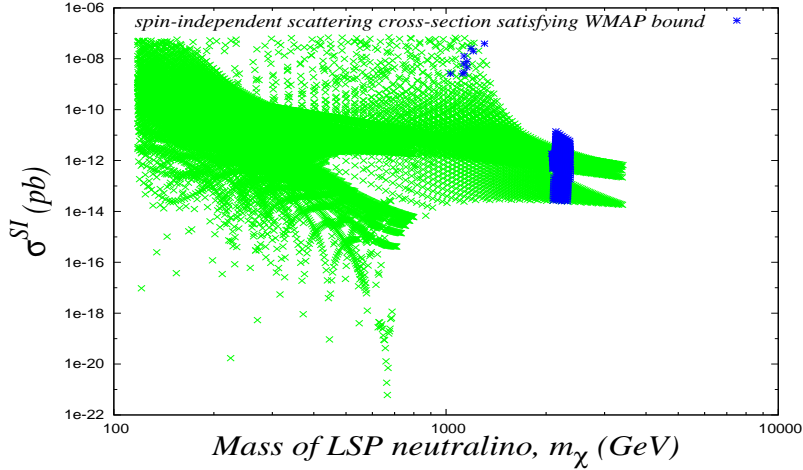}
\caption{\label{fig:si} \textit{The plot showing the variation of
Spin Independent scattering cross
section ($\sigma^{SI}$) with for Mass of the LSP neutralino ($m_{\chi}$) for the allowed
SUSY parameter space. The blue zones are the LSP neutralinos which satisfy the WMAP relic.}}
\end{center}
\end{figure}

\begin{figure}[h]
\begin{center}
\includegraphics[width=2.5in,height=4.2in,angle=-90]{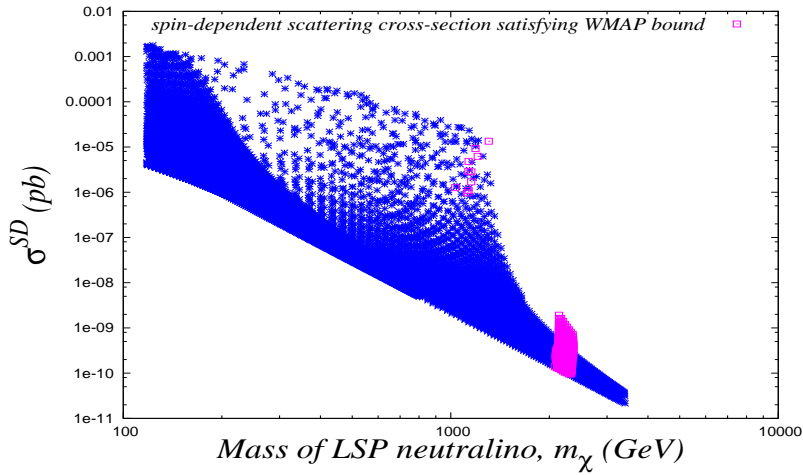}
\caption{\label{fig:sd} \textit{The variation of Spin Dependent scattering cross
section ($\sigma^{SD}$) with mass of the LSP neutralino 
($m_{\chi}$) allowed parameter space is shown in this plot. WMAP relic satisfied
two zones are shown in pink.
}}
\end{center}
\end{figure}

We have computed both spin-independent and spin-dependent 
scattering cross sections of neutralino dark matter
for a wide range of mass in this model respecting the
allowed $m_0 - m_{3/2}$ 
bound. As the nucleon consists 
of both protons and neutrons, the WIMPs can be scattered
off both nucleons. The contribution of 
loop diagrams along with the
tree level diagrams have also been included for calculations 
of scattering amplitudes for 
both SI and SD cases of $\chi\rm-N$ scattering.
These scattering cross sections for different neutralino masses 
are computed using \texttt{micrOMEGAs} \cite{micromegas}
computer code. The results for both SI and SD cases are  
shown in Fig. \ref{fig:si}
and Fig. \ref{fig:sd} respectively as scattered plots for
scattering with protons. 

The green scattered plots in Fig. \ref{fig:si} give the 
spin independent scattering cross section $\sigma^{\rm SI}$ 
for various neutralino masses $m_\chi$ generated in the present 
AMSB model with the bound on 
parameter space.
The blue scattered plots in Fig. \ref{fig:sd}, on the other 
hand, are for the spin dependent case. 
The mass region(s) in this model that satisfy the WMAP results 
for relic density are superimposed over these two figures 
in order to constrain the $m_\chi - \sigma^{\rm SI/SD}$ space 
obtained from Figs. \ref{fig:si}, \ref{fig:sd}. The blue patches in 
Fig. \ref{fig:si} and the pink patches in Fig. \ref{fig:sd} represent
the mass regions that satisfy WMAP results. 
Clearly, there are two different zones allowed by the WMAP
limits as expected from the discussions in Sect. 2. 
For the WMAP allowed lower mass region 
(around 1 TeV), the SI cross section (Fig. \ref{fig:si}) extends between
$\sim 10^{-9}\rm{-}\sim 10^{-7}$ pb. The WMAP allowed higher mass region 
(around 2 TeV) which spans larger region in $m_\chi - \sigma^{\rm SI}$
space than the WMAP allowed lower mass region, is confined within SI
cross section limit
$\sim 10^{-11}\rm{-}\sim 10^{-14}$ pb in Fig. \ref{fig:si}. 
The pink regions in
Fig. \ref{fig:sd} signify the WMAP allowed region. Here, the value of 
SD scattering cross section is coming to be higher
than that of SI as it is expected from the theoretical perspective. 

\begin{figure}[h!]
\begin{center}
\subfigure{
\includegraphics[width=3.0in,height=2.0in]{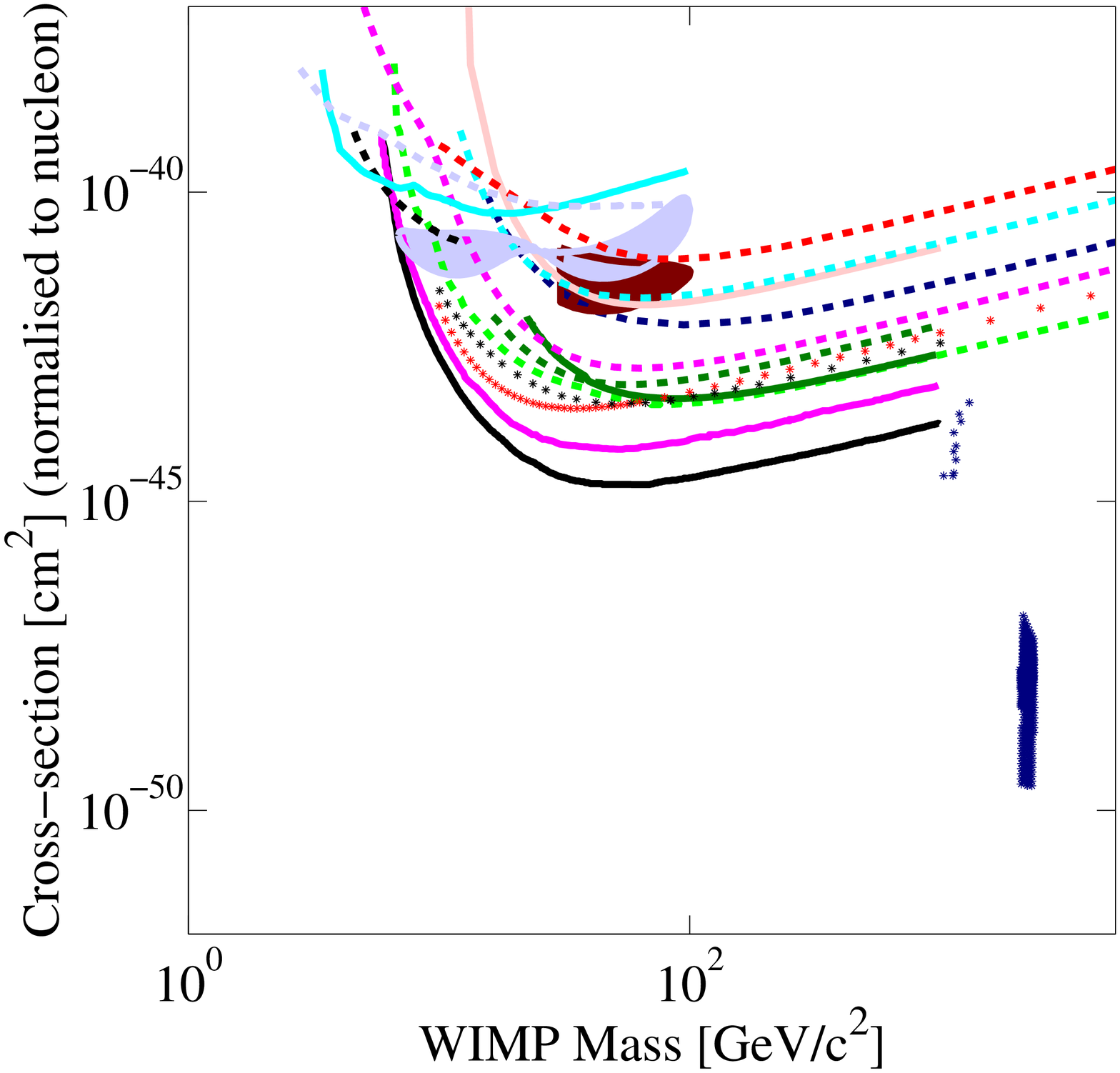}}
\subfigure{
\includegraphics[width=2.4in,height=2.0in]{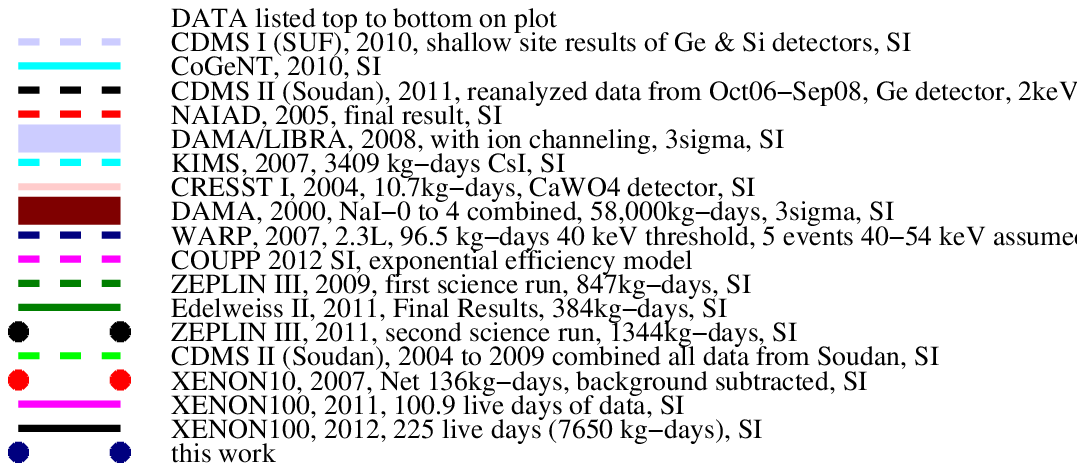}}
\caption{\label{fig:silimit} \textit{Limits on spin independent 
scattering cross sections set by various experiments and comparison with our
results in mAMSB model. Our calculated results that
follow the WMAP limits are shown by two distinct blue patches in this figure
and they are found to be within these experimental bounds.} }
\end{center}
\end{figure}

Similarly, in Fig. \ref{fig:sd}
the WMAP allowed lower mass region (around 1 TeV) constrain the spin dependent 
cross section $\sigma^{\rm SD}$ limits in the range
$\sim 10^{-6}\rm{-}\sim 10^{-5}$ pb and for the region of around 
2 TeV $\sigma^{\rm SD}$ lies between $\sim 10^{-10}$ to $\sim 10^{-9}$ pb.
We mentioned in passing that we obtained similar nature for 
WIMP-{\it neutron} elastic scattering. 

\begin{figure}[h!]
\begin{center}
\subfigure{
\includegraphics[width=3.0in,height=2.0in]{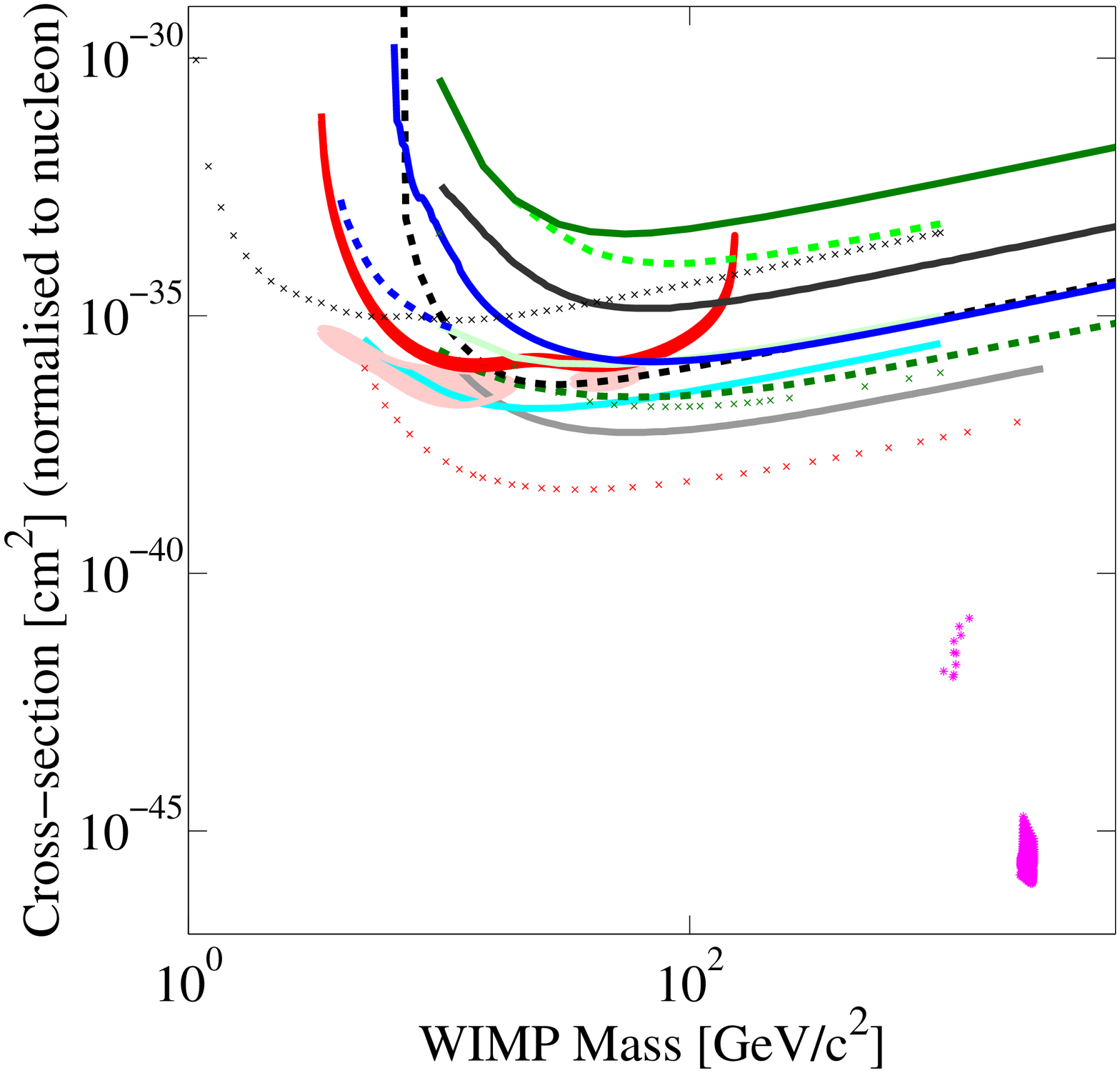}}
\subfigure{
\includegraphics[width=2.4in,height=2.1in]{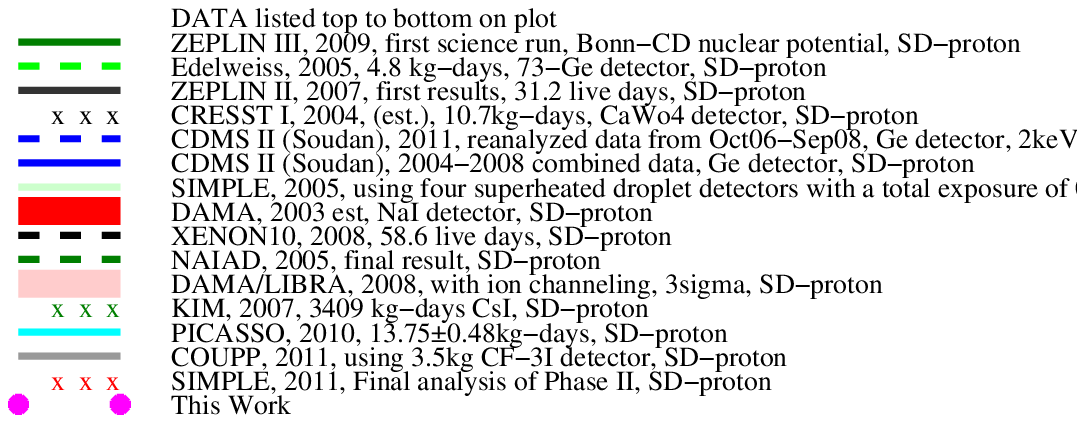}}
\caption{\label{fig:sdlimit} \textit{Limits on spin dependent scattering 
cross sections set by various experiments and comparison with our calculated
results. The pink zones are the satisfying WMAP bounds and they are few orders
below the upper bounds of these experimental data.} }
\end{center}
\end{figure}

Figs. \ref{fig:silimit} and \ref{fig:sdlimit} 
show respectively
various upper limits in dark matter mass - SI or SD scattering cross section
($m_\chi - \sigma^{{\rm SI}\,{\rm or}\,{\rm SD}}$) plane 
set by different ongoing direct detection experiments.
The WMAP-allowed regions from the present model for 
neutralino dark matter are superimposed on them for comparison.
The experimental limits are obtained from the compilation given in 
Ref. \cite{dmlimitplot}. The names of the different experiments 
are furnished as legends in the figures. It is obvious from 
Figs. \ref{fig:silimit} and \ref{fig:sdlimit} that the allowed parameter
space for the considered AMSB model is within the allowed limits
of the experimental bounds.

\section{Indirect Detection}

Weakly interacting dark matter in our galaxy can
be trapped inside massive heavenly bodies like galactic centre or the sun 
due to the gravity of these bodies. The dark matter particles, in course of 
their passage through such massive bodies undergo elastic 
scattering off the nuclei present there as a result of which  
their velocity deplete. If their velocities fall below their 
escape velocities from such massive objects, the dark matter 
particles are trapped. These trapped dark matter particle ($\chi$) may undergo 
the process of pair-annihilation producing primarily $b$, $c$ and $t$ 
quarks, $\tau$ leptons, gauge bosons, etc. 
($\chi\chi\rightarrow q\bar{q}, l^+l^-, \nu\bar{\nu}, ZZ, W^+W^-,\,...$).
The annihilation products depend on the mass and composition of the 
dark matter. Neutrinos and antineutrinos can be produced 
by the decay of primary annihilation products or through 
direct annihilation.

The main principle of indirect detection of dark matter is to 
detect and measure the fluxes of standard model particles produced 
from the annihilation of dark matter trapped by the gravitation of 
massive heavenly bodies. 
Recently many new results from indirect 
DM searches have been released. An interpretation of these 
excesses related to astrophysical processes from any galactic or
extragalactic sources is still not very clear. 
The products from the annihilation of dark matter particles in 
massive bodies such as in galactic centre may explain 
such excess signals. 

There are a lot of satellite borne experiments that look for 
gamma rays or antimatters in cosmos. Some terrestrial experiments
are also suited for looking at cosmic gamma rays, neutrinos etc. 
Such experiments include 
PAMELA \cite{pamela} that confirms an excess in positron 
fraction in agreement 
with earlier indications by HEAT \cite{heat} and AMS01 \cite{ams01}. 
Other satellite borne 
experiments like FERMI \cite{fermie} and ATIC \cite{atic} report an
excess in total electron and positron spectrum at energies of several hundreds
of GeV's, much higher than that of PAMELA search.
The cosmic gamma rays from the galactic sources and from galactic centre 
are measured in a wide range of energies by INTEGRAL
($<$ $\sim$1 MeV) \cite{integral}, EGRET \cite{egret}, FERMI \cite{fermi}, HESS \cite{hess1,hess2},
MAGIC \cite{magic}, Whipple/Veritas \cite{whipple}, CANGAROO ($>$ $\sim$100 GeV)
\cite{cangaroo} etc.

In this work we mainly focus on the gamma ray and neutrinos from
dark matter annihilations in the direction 
at and around galactic centre (GC). The GC region has  
higher dark matter density and hence  
a promising site for the study of indirect detection of dark matter.
Although GC seems to be the most obvious
target, it is also
one of the most difficult areas to work with because of the complex and
poorly-understood backgrounds \cite{vitale,acero}, 
for signals from around GC 
and uncertain dark matter profile \cite{stoehr,merritt,varitas}. 

The galactic gravitational potential leads to a higher dark matter density 
at the centre of Milky Way. The expected flux from the
galactic centre depends on the distribution of dark matter in the galaxy. 
The dark matter density profile $\rho(r)$ is assumed
to be spherically symmetric. 
The differential flux of the outgoing particle of type $i$ is given by
\begin{equation}
 I^i(E,\theta) = \frac{d\Phi_{i}}{dE} = 
\sum_j\frac{\sigma_j\upsilon}{8\pi\alpha m_{\chi}^2}
\frac{dN^i_j}{dE}(E)J(\theta,\Delta\Omega)
 \label{flux}
\end{equation}
where the factor, $\alpha$ is 1 or 2 depending on whether the assumed WIMPs 
are self-conjugated or not respectively. In the above `j' denotes a particular 
annihilation channel. It is also to be mentioned that 
the effect of this factor, $\alpha$ on the above differential 
flux is much less significant 
in comparison to the dark matter 
density fluctuations in the innermost regions of
Milky Way. Here we consider $\alpha$ to be unity as the 
neutralinos from 
the mAMSB model (the dark matter candidate chosen in the present work)
are self-conjugated. In Eq. \ref{flux}, 
$\sigma$ is the annihilation cross section of dark matter and
$\upsilon$ denotes the relative velocity of the dark matter particles. 
The quantity
$\frac{dN^i}{dE}(E)$ in Eq. \ref{flux} is the energy spectrum
of particle $i$ and $J(\theta,\Delta\Omega)$ is given by,
\begin{equation}
 J(\theta,\Delta\Omega) = \int_{\Delta \Omega}{d\Omega} 
\int_{\rm line\,of\,sight}{\langle\rho^2(r(\tilde r,\theta))\rangle d\tilde r}\,\, .
\label{jomega}
\end{equation}
With $\theta$ being the angle subtended by the line of sight of an 
observer on the earth (along the length $\tilde r$) on $R_\odot$ $-$
the distance between GC and the terrestrial observer (in solar system).
The source to observer distance $\tilde r$ can be calulated as 
\bea
\tilde r &=& \sqrt{(r^2+R_\odot^2-2rR_\odot cos\theta)}\,\, ,
\eea
In the above, the target region is considered to  
be at a distance $r$ from GC (at the GC, $r$ = 0).   
Here we also mention that the GC
is assumed to be coincident with the halo centre).
The solar system's position 
in the halo from the GC is given by $R_\odot$ = 8.0 kpc.  
In Eq. \ref{jomega}, $\Delta \Omega$ is the solid angle over 
which the observation is to be made and $\rho(r)$ is the dark matter 
density at a distance $r$ from GC. Clearly the integration on the 
RHS of Eq. \ref{jomega} is along the line of sight. Thus the 
astrophysical factor $J$ in Eq. \ref{flux} has only a $\theta$ 
dependence (along with $\Delta \Omega$) and thus the differential flux 
$I_{\gamma}$ can be expressed in terms of the angle $\theta$
corresponding to different positions of the source in galactic halo
with respect to GC.  
 
\begin{figure}[h]
\begin{center}
\includegraphics[width=2.7in,height=4.5in,angle=-90]{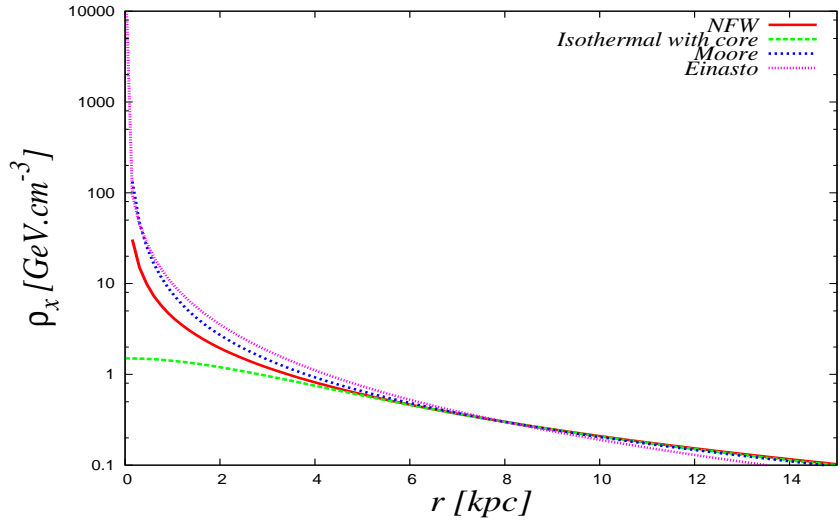}
\caption{\label{fig:halo} \textit{The variation of galactic halo density with
radial distance for various halo models and the cuspy or flat nature of
the considered halo profiles are shown.}}
\end{center}
\end{figure}

The dark matter density $\rho(r)$ is related to   
the spherically symmetric 
halo profile of galactic dark matter by the equation 
\bea
\rho(r) &=& \rho_0F_{\rm halo}(r)\,\, ,
\label{rho}
\eea
where $\rho_0$ is the dark matter density at 
the galactic centre assumed to be 
0.3 GeV/cm$^3$
and $F_{\rm halo}(r)$ is the halo profile of the galactic
dark matter which can be expressed in a parametric form,
\bea
F_{\rm halo}(r) &=& \left[\frac{R_\odot}{r}\right]^\gamma 
\left[\frac{1+\left[\frac{R_\odot}{a}\right]^\alpha}
{1+\left[\frac{r}{a}\right]^\alpha}\right]^{\frac{\beta-\gamma}{\alpha}}\,\,.
\label{fhalo}
\eea
In the above, $a$ is a scale parameter and 
the other parameters $\alpha$, $\beta$, $\gamma$  
take different values 
for different halo models which follow the above parametric form
for $F_{\rm halo}$. For example,
for NFW halo profile \cite{nfw}, $\alpha=1$, $\beta=3$, $\gamma=1$ 
and $a=20$ kpc, whereas the parameter set
$\alpha=2$, $\beta=2$, $\gamma=0$ and $a=4$ kpc represents 
isothermal profile with core \cite{iso}. Again for the 
Moore profile \cite{mre} we have,
$\alpha=1.5$, $\beta=3$, $\gamma=1.5$ and $a=28$ kpc.
In Einasto halo profile \cite{ein} however, a different 
kind of parametric form is adopted   
which is given by,
\begin{equation}
 F^{Ein}_{\rm halo}(r)=exp\left[\frac{-2}{\tilde \alpha}
\left(\left(\frac{r}{R_\odot}\right)^{\tilde \alpha}-1\right)\right]\,\,\, ,
\end{equation}
where $\tilde \alpha$ is the parameter. In this work   
$\tilde \alpha = 0.17$ is adopted. In what follows 
the four profiles are referred to as NFW, Isothermal, 
Moore and Einasto respectively. The galactic halo densities 
for these four halo models are shown in Fig. \ref{fig:halo}.
In the present work we show the gamma ray and neutrino flux 
from galactic centre region for each of the four halo profiles 
mentioned above.

\subsection{Gamma Ray Flux Results}

\begin{figure}[h!]
\begin{center}
\subfigure[NFW\label{fig:gnfw}]{
\includegraphics[width=2.2in,height=2.9in,angle=-90]{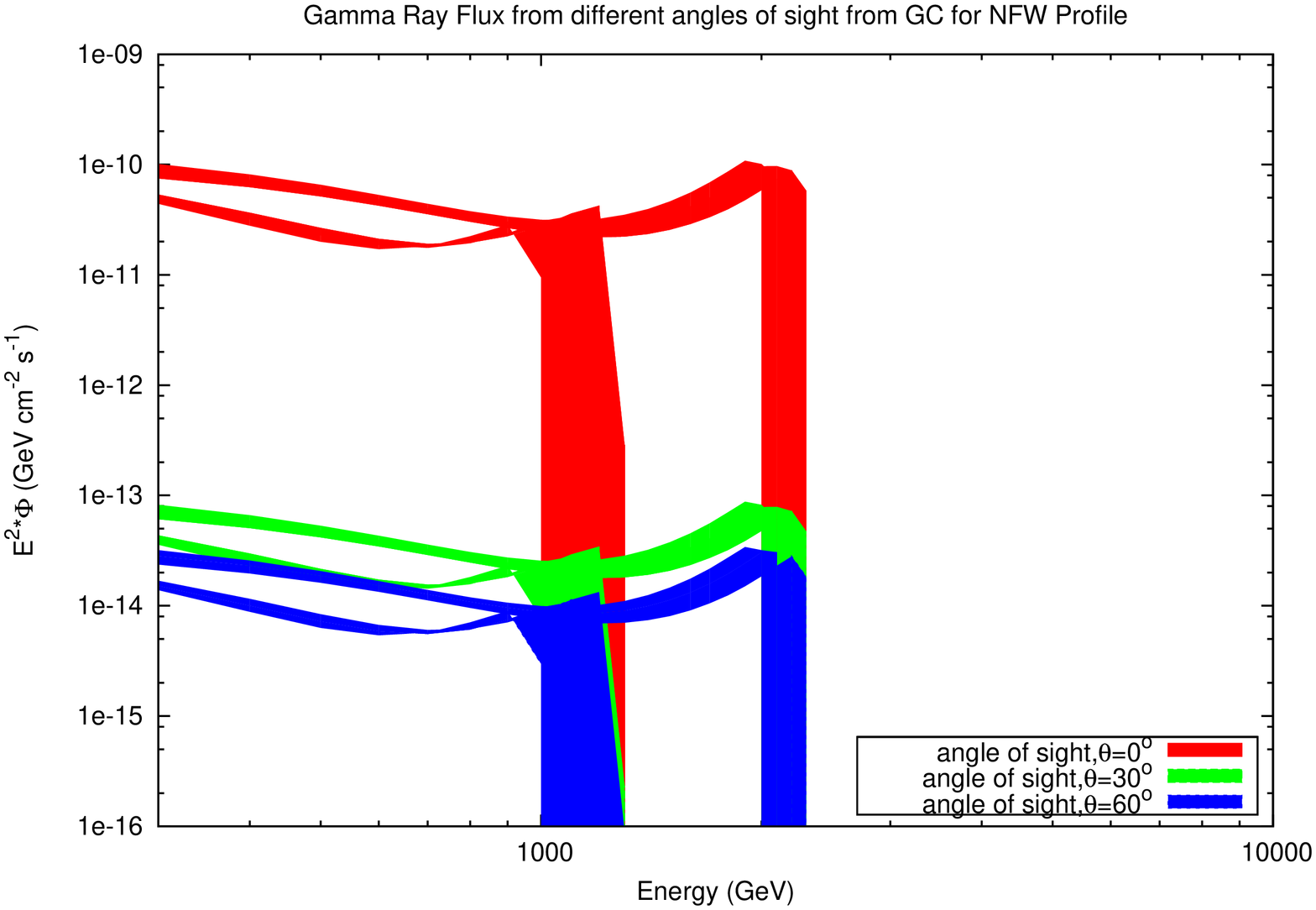}}
\subfigure[Isothermal\label{fig:giso}]{
\includegraphics[width=2.2in,height=2.9in,angle=-90]{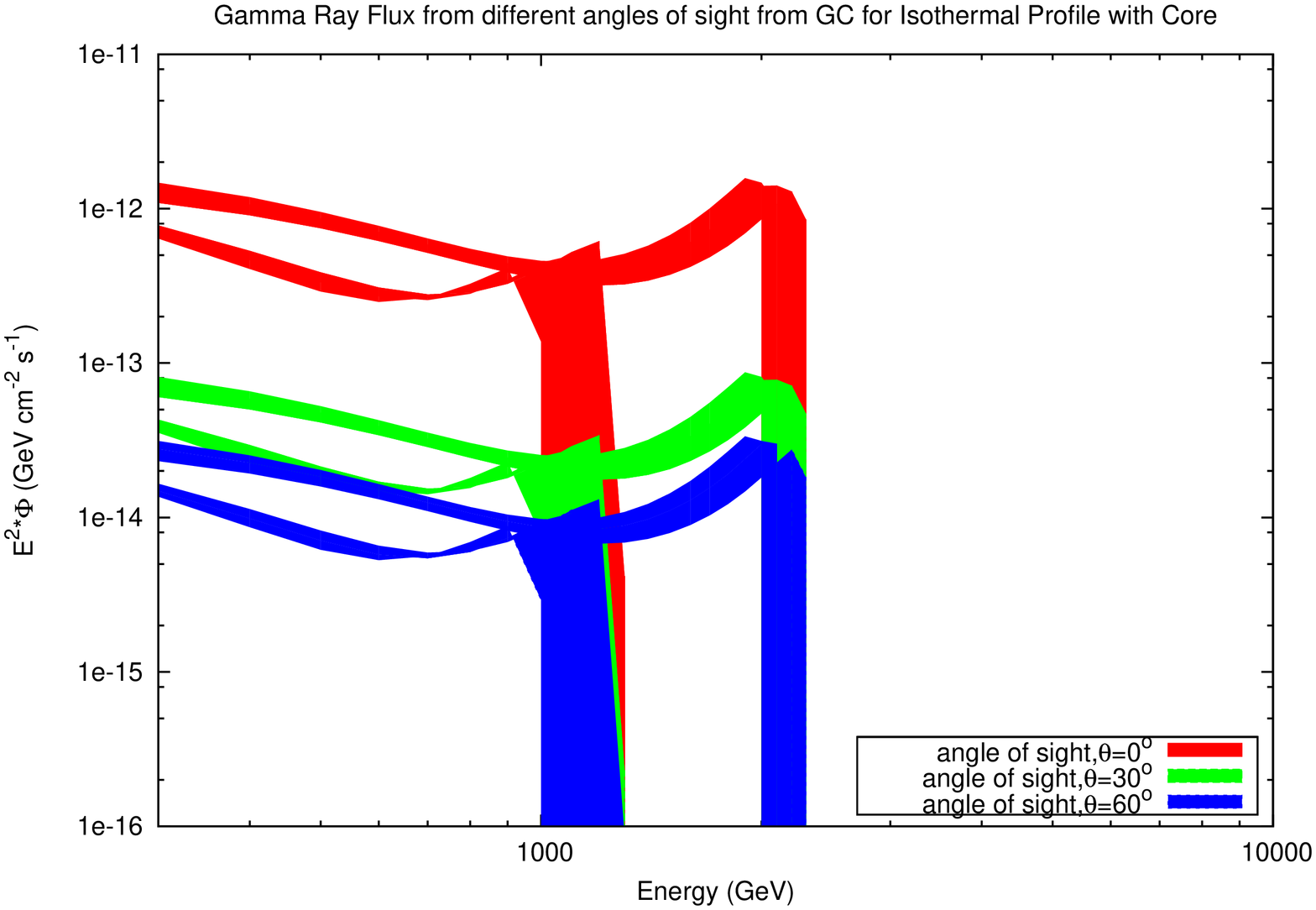}}
\subfigure[Moore\label{fig:gmre}]{
\includegraphics[width=2.2in,height=2.9in,angle=-90]{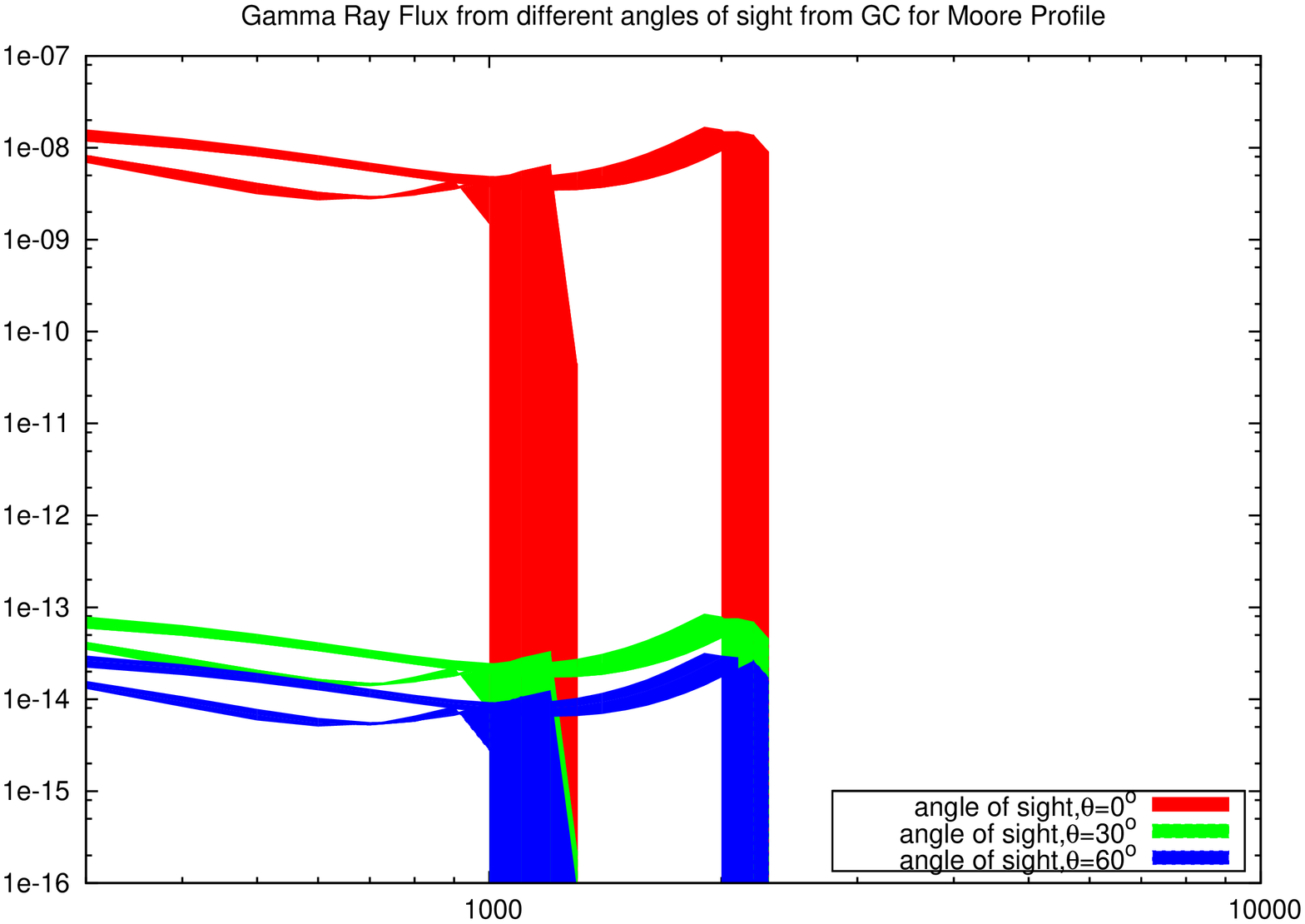}}
\subfigure[Einesto\label{fig:gein}]{
\includegraphics[width=2.2in,height=2.9in,angle=-90]{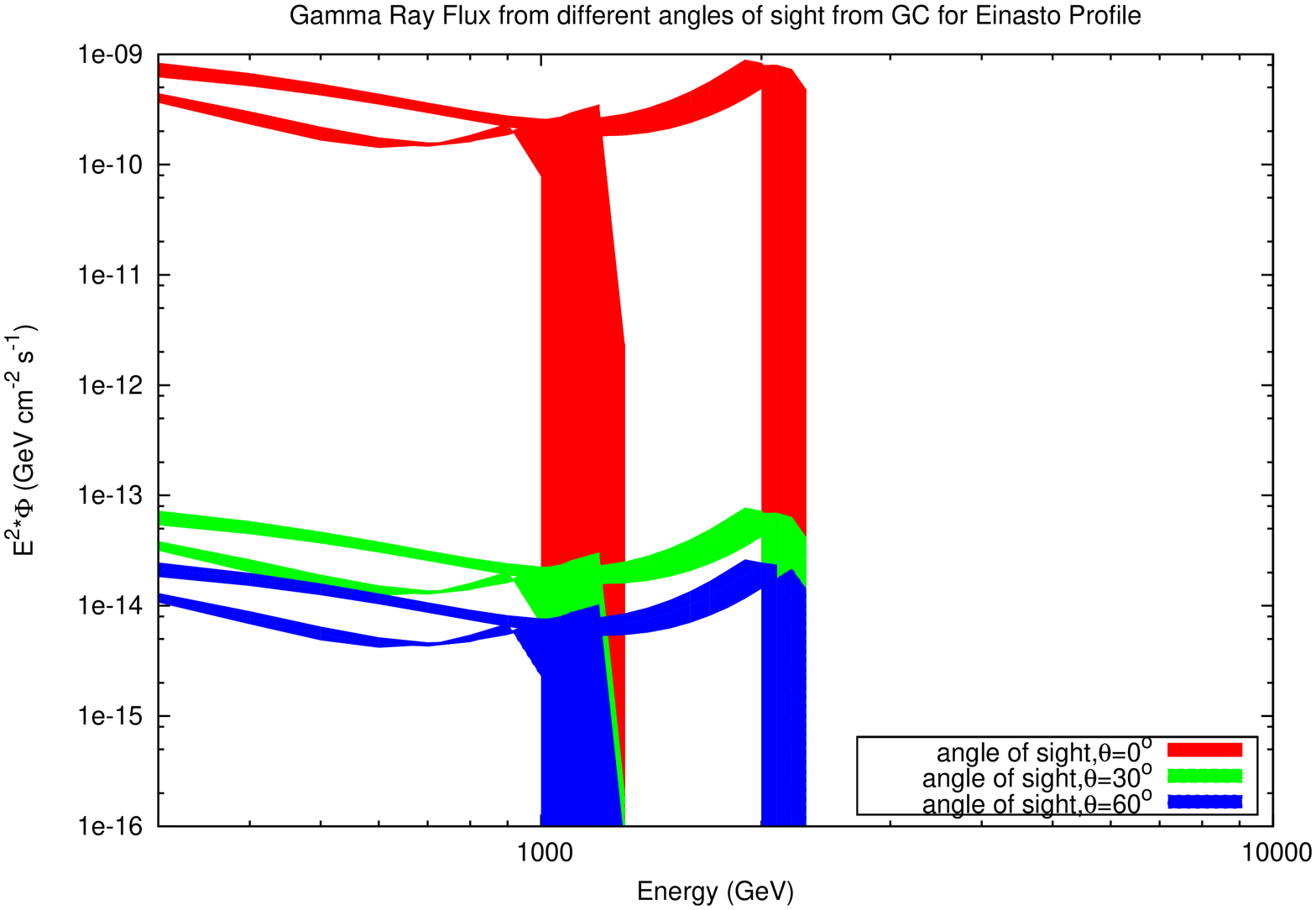}}
\caption{\label{fig:gammaflux} \textit{ Plot showing the  
variation of gamma ray flux with energies from the annihilation of 
dark matter for 
different galactic DM halo models and for different angles of sight, $\theta$.
The red lines describe the flux observed at $\theta = 0^o$, i.e.,
from the galactic centre and the green and blue coloured
regions are for the observations at $\theta$ = $30^o$ and $60^o$
respectively. The subfigures are for different commonly used
dark matter halo profiles implemented in this work,
a) NFW profile b) Isothermal profile with core
c) Moore profile d) Einasto profile}}
\end{center}
\end{figure}

There are generally two kinds of $\gamma$-ray emission from 
DM annihilation. In the first category $\gamma$ is produced
directly from the annihilation final state particles which is called 
primary emission in which final charged leptons
emit gamma ray or $\pi^0$ which eventually decays to gamma ray after hadronization. 
The other kind is called secondary emission in which
gamma rays are produced by interactions of final state particles with 
external medium or radiation field such as the inverse Compton effects etc. 
Here we consider only the first type of emission for which
the relation (\ref{flux}) holds. 
Here we calculate the gamma ray 
flux from the galactic centre as also from other places 
in galactic dark matter halo along the line of sight around the GC. As discussed 
in the previous section, the targets away from the galactic
centre are characterised by changing only the angle $\theta$. 
This angle $\theta$ in fact denotes the angle of sight from 
the observer with respect to the line of 
sight when the observer is looking directly at the galactic centre.
The polarisation effect of final state gauge bosons 
($W^{\pm}$ and $Z$)
and also the photon radiation effect which
strongly affect the gamma ray spectra are also taken into account in the 
present work. The $\gamma$-flux is computed using \texttt{micrOMEGAs} code. 
The calculations are made for each the four halo profiles,
referred to as NFW, Isothermal, Moore and Einasto and the 
results are furnished in the four figures namely
Figs. \ref{fig:gammaflux}a - \ref{fig:gammaflux}d
respectively.   

In Figs. \ref{fig:gammaflux}a - \ref{fig:gammaflux}d, we plot the 
quantity $E^2 \times \frac {d\Phi}{dE}$ for different values 
$E$, the energy of the emitted $\gamma$ rays from dark matter annihilations. 
We show the results 
for the cases when $\theta =0$ (galactic centre), 
$\theta = 30^o$ and $\theta = 60^o$ and are shown as red, green and blue 
regions respectively. One notices in  
Figs. \ref{fig:gammaflux}a - \ref{fig:gammaflux}d   
that $\gamma$ flux for any particular value of the angle $\theta$ 
are given as a pair of plots designated by the specific colour code 
(red, blue or green), assumed for the results
corresponding to that particular $\theta$. This is due to 
the fact that the WMAP data constrain the supersymmetric 
parameter space considered in this work, in two distinct zones as shown in 
Figs. \ref{fig:relic}, \ref{paramscan} and each of the plots 
in every such pair of $\gamma$ flux in Figs. \ref{fig:gammaflux}a - 
\ref{fig:gammaflux}d  
correspond to each of the WMAP 
allowed regions for the present AMSB model for cold dark matter candidates.   
It is clear from Figs. \ref{fig:gammaflux}a - \ref{fig:gammaflux}d
that calculations with different halo profiles yield different results for 
$\gamma$ flux. It is also to be noted that the flux in the direction of the galactic 
centre ($\theta=0$) is larger than the flux from other directions 
(corresponding to different values of $\theta \neq 0$)
for each of the four halo models considered.  
    
The $\gamma$ fluxes are found to be 
almost of the similar order for the cases when
$\theta = 30^o$ and $\theta = 60^o$
in each of the Figs. \ref{fig:gammaflux}a - \ref{fig:gammaflux}d.
This reflects the fact that the
DM halo profiles are almost flat in those regions.
The Einasto profile has a finite (zero) central slope
unlike the NFW profile which has a divergent (infinite) central density.
As it is not yet known which model provides
the best description of the central densities of simulated dark-matter halos,
we have taken these known models into
account.

\begin{figure}[h!]
\begin{center}
\subfigure[\label{fig:hessnfwa}]{
\includegraphics[width=2.2in,height=2.9in,angle=-90]{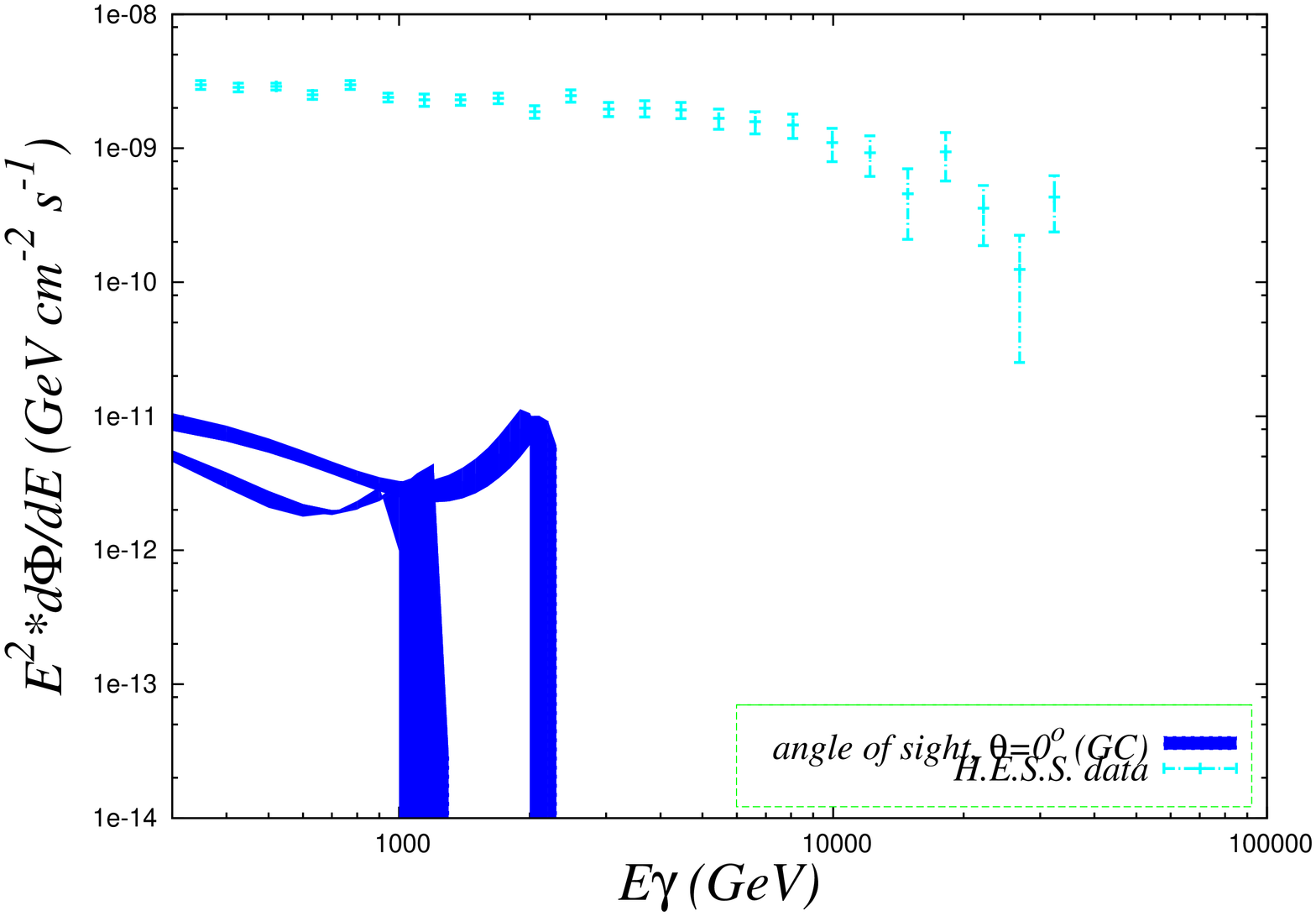}}
\subfigure[\label{fig:hessnfwb}]{
\includegraphics[width=2.2in,height=2.9in,angle=-90]{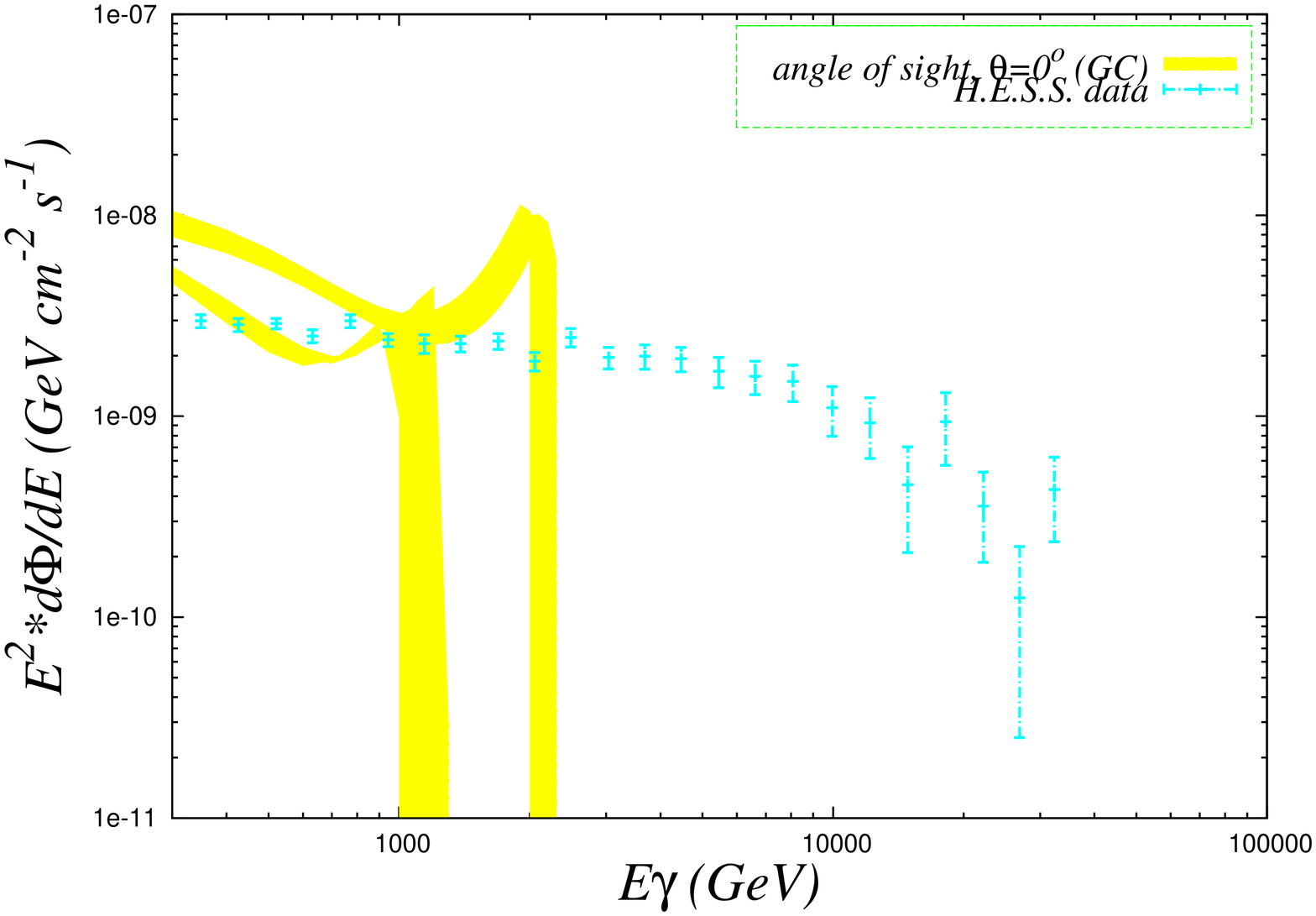}}
\caption{\label{hessnfw} \textit{Plot of energy vs. $\gamma$ flux
for dark matter annihilation at the galactic centre
and comparison with the HESS experimental data
for NFW profile a) without baryonic compression and b) with
the baryonic compression and $\sim 10^3$ flux enhancement }}
\end{center}
\end{figure}

\begin{figure}[h!]
\begin{center}
\subfigure[\label{fig:hesshaloa}]{
\includegraphics[width=2.2in,height=2.9in,angle=-90]{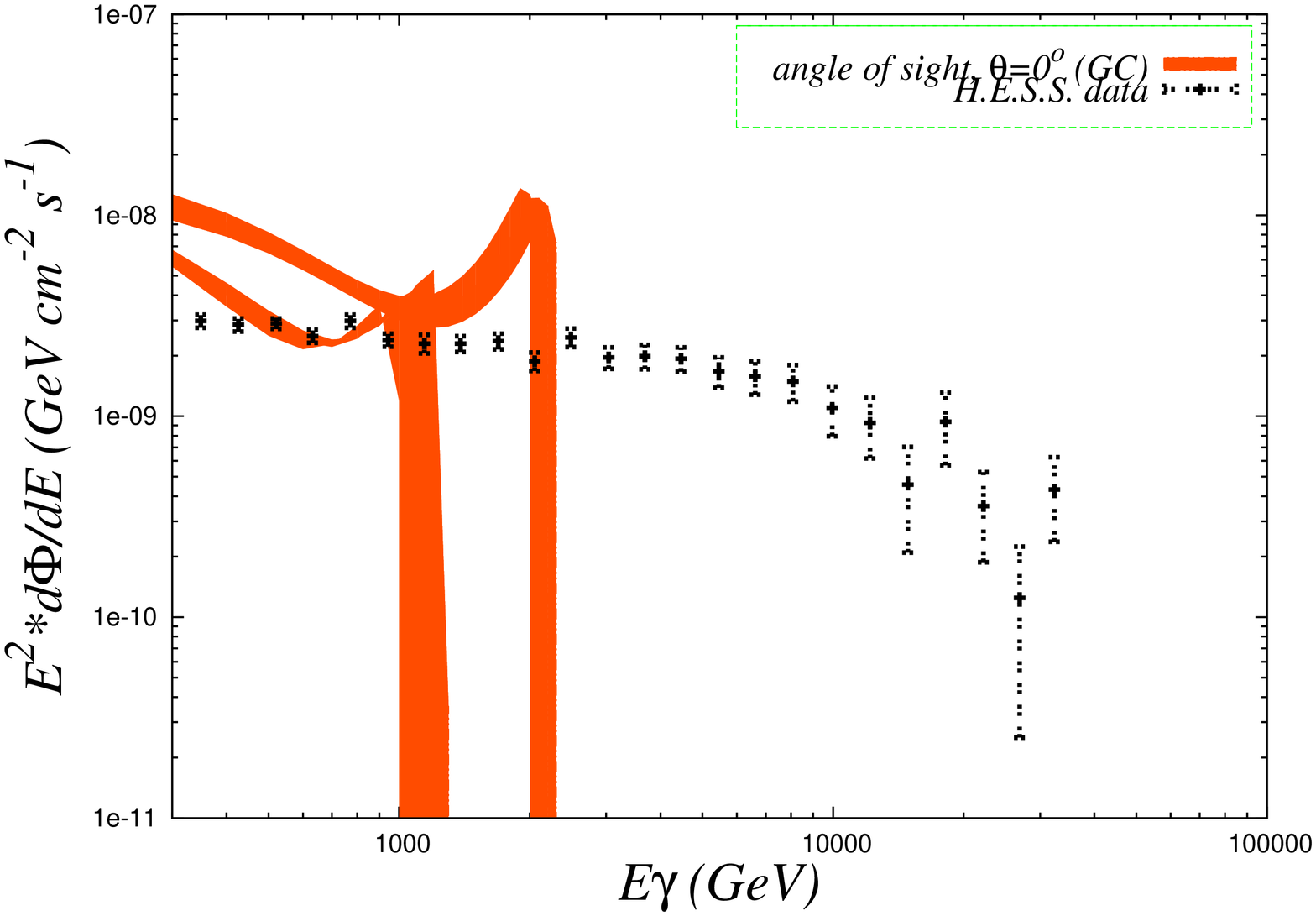}}
\subfigure[\label{fig:hesshalob}]{
\includegraphics[width=2.2in,height=2.9in,angle=-90]{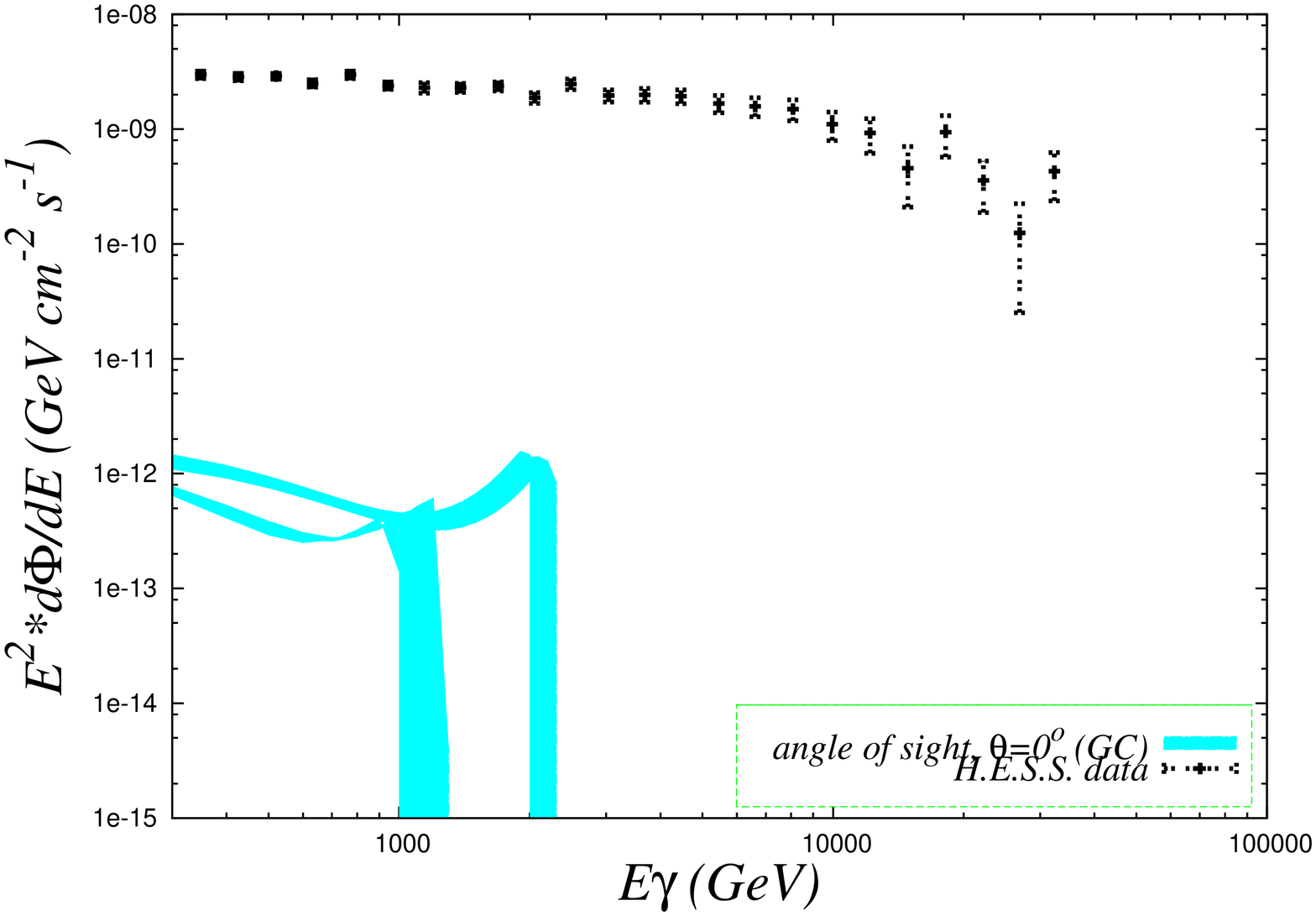}}
\subfigure[\label{fig:hesshaloc}]{
\includegraphics[width=2.2in,height=2.9in,angle=-90]{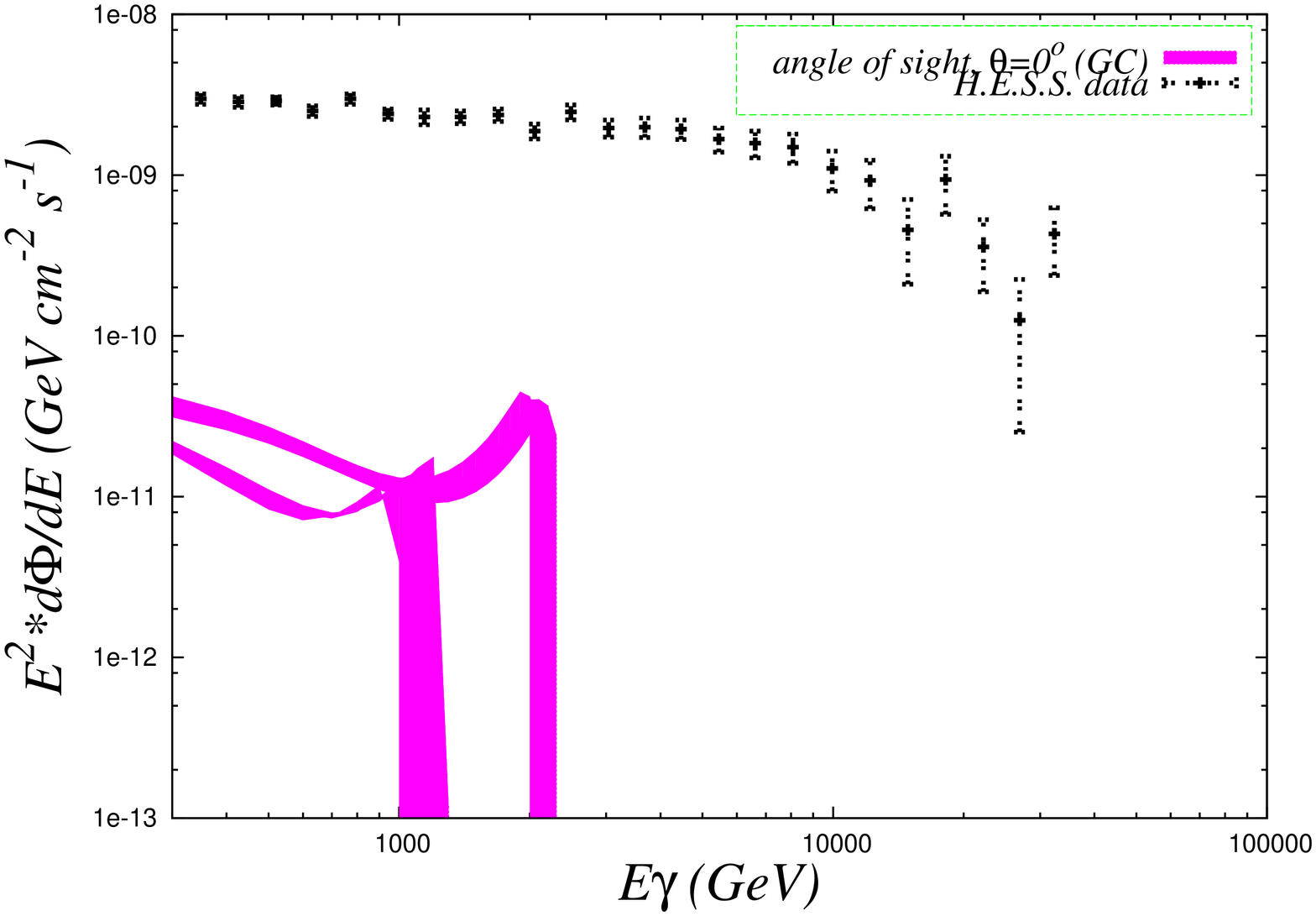}}
\caption{\label{hesshalo} \textit{Plot of energy vs. $\gamma$ flux 
from DM annihilation at the GC and comparison
with the HESS experimental data for a) Moore
profile, b) Isothermal profile with core and c) Einasto profile}}
\end{center}
\end{figure}

The $\gamma$-flux thus obtained for different 
halo models are compared with the 
observational results of The High Energy Stereoscopic System (HESS)
experiment. Located in Namibia, the HESS experiment is designed to investigate 
high energy cosmic gamma rays ($\sim 100$ GeV - TeV energy ranges) and it can
also investigate the $\gamma$-rays  
in its observable energy range which can be due to the annihilation
of cold dark matter particles. The results are given in 
Fig. \ref{hessnfw} and Fig. \ref{hesshalo}. In each of the figures, 
the calculated flux are shown by two diferent regions corresponding to 
WMAP constrained two zones of dark matter mass in the present 
mAMSB model (discussed earlier).  
It has been argued 
by Prada {\it et al.} in Ref. \cite{boostprada} (and also
in Ref. \cite{mambrini}) that due to the infall of 
baryons at the galactic centre, the expected $\gamma$ signal 
from dark matter annihilation at galactic centre will be 
boosted in case the dark matter consists of supersymmetric particles. 
In fact, considering neutralino in minimal supergravity (mSUGRA) model
as the candidate for dark matter and with the NFW dark matter halo
profile, they have demonstrated that the 
said boost can be of the order of 1000. In Fig. \ref{hessnfw}a,
the $\gamma$ flux from the galactic centre as calculated from the 
annihilation of neutralino dark matter in present mAMSB model 
assuming the NFW profile, is compared with the HESS results. 
The solid angle at which the HESS experiment looks at the 
galactic centre is $\sim 10^{-5}$ sr, a value which is also 
adopted in the present calculations to obtain the 
results shown in Fig. \ref{hessnfw} and Fig. \ref{hesshalo}.
It is evident from Fig. \ref{hessnfw}a that the 
$\gamma$-flux obtained from the present calculations 
is much less than the HESS results for the energy
range given by the model with WMAP constraints. 
In Fig. \ref{hessnfw}b we show a representative plot where 
the calculated $\gamma$-flux is multiplied (``boosted") by a factor
of 1000 and then compared with the HESS results. Fig. \ref{hessnfw}b
shows that the ``boosted" flux is in the similar ball park of 
HESS results which seems to satisfy the claim made in Ref. \cite{boostprada}.  
In Figs. \ref{hesshalo}a, \ref{hesshalo}b and \ref{hesshalo}c, 
we show similar comparisons with HESS results for calculations made 
with Einasto, Moore and isothermal halo profiles respectively. 
One observes from Fig. \ref{hesshalo} that both for isothermal 
and Einasto profiles, the calculated fluxes are below the HESS results
while for Moore profile, they are comparable with HESS results. 
Both NFW and Moore profiles are cuspy in nature and they essentially differ 
by the values of the parameters $\alpha$, $\beta$, $\gamma$.
On the other hand both the Einasto and isothermal profiles are 
non-cuspy in nature while the latter is a flat halo profile. One  
needs to increase the calculated flux by a factor $\sim 10^{2}$ for 
the former case while the calculated flux for the isothermal profile 
needs a boost of $\sim 10^{5}$ to be in the regime of HESS observational
results.   

\subsection{Neutrino Flux Results}

As discussed earlier, neutrinos can also be produced 
by the annihilation of two neutralinos --
the present dark matter candidate.  
These trapped dark matter at the galactic centre  
produces primarily $b$, $c$, $t$ quarks, 
$\tau$ leptons, gauge bosons, etc. through the process of pair-annihilations. 
The neutrinos can be obtained from the decay or pair annihilation 
of the primary products. The neutrinos can also be produced
directly from the annihilation of two mAMSB neutralinos ($\chi \tilde{\chi}
\rightarrow \nu \bar{\nu}$) mediated by $Z$, sneutrino 
($\tilde{\nu}$) etc. 
In this work we investigate 
the muon neutrino ($\nu_\mu$) flux from the galactic centre 
due to the annihilation of such neutralinos in the present mAMSB model
and its possible detection prospect at an earthbound detector. 
Searches for neutralino annihilation into neutrinos is subject to 
extensive experimental
investigations in view of the neutrino telescopes like
IceCube \cite{icecube}, Baikal \cite{baikal},
NESTOR \cite{nestor}, ANTARES \cite{antares}. 
The calculation of flux of
neutrinos coming from GC are similar to that of gamma rays as 
both are electromagnetically neutral particles. So, they are
not affected by the irregularities of galactic magnetic 
fields or any magnetic turbulences. Also, they do
not suffer any energy loss from inverse compton 
effect or from synchrotron radiation.
For the present 
case we calcuate the possible muon ($\mu$) signal from these neutrinos at ANTARES 
neutrino telescope \cite{antares} installed in the sea-bed off France coast. 

\begin{figure}[h!]
\begin{center}
\subfigure[\label{fig:nfw1}]{
\includegraphics[width=2.2in,height=2.9in,angle=-90]{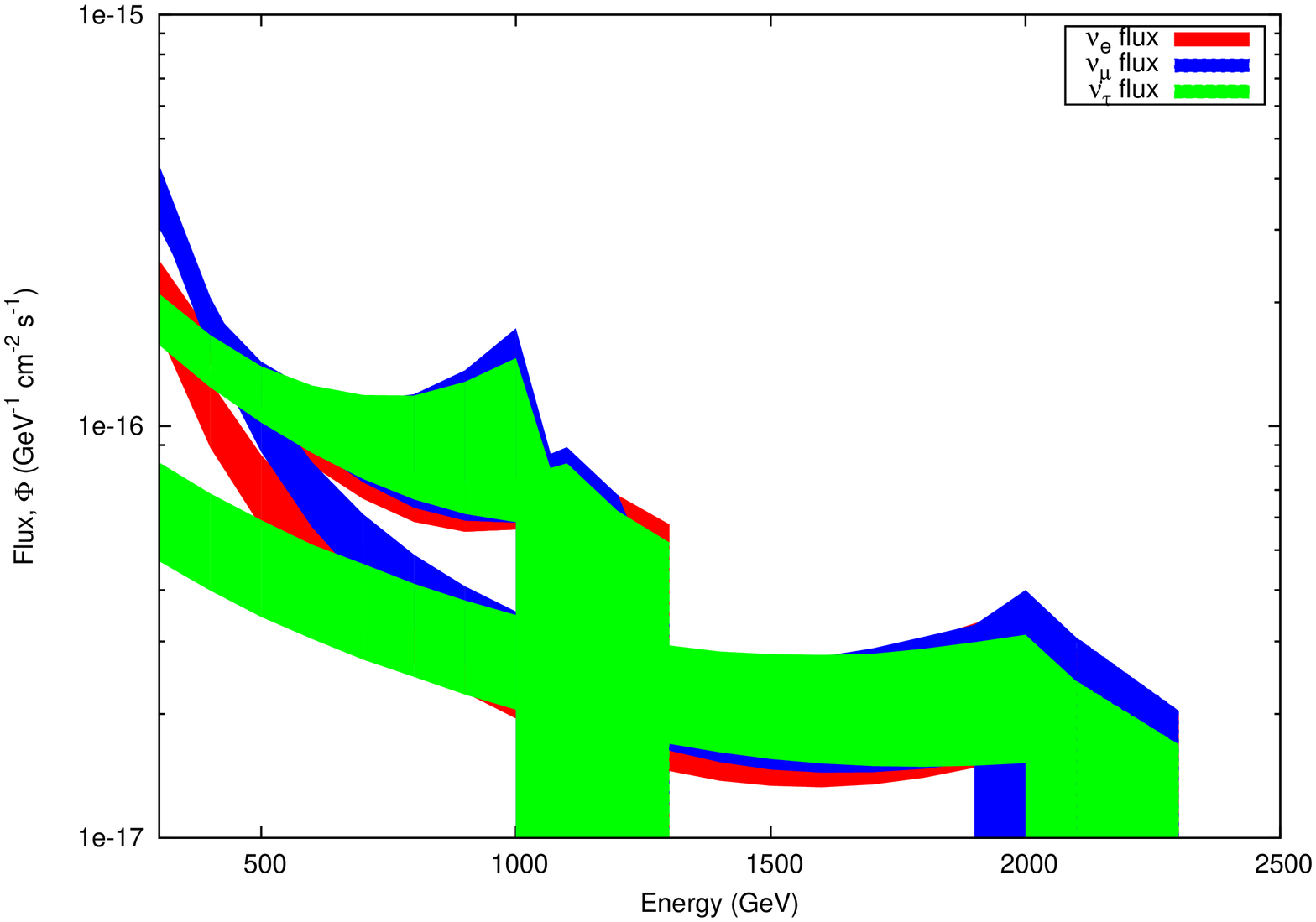}}
\subfigure[\label{fig:nfw2}]{
\includegraphics[width=2.2in,height=2.9in,angle=-90]{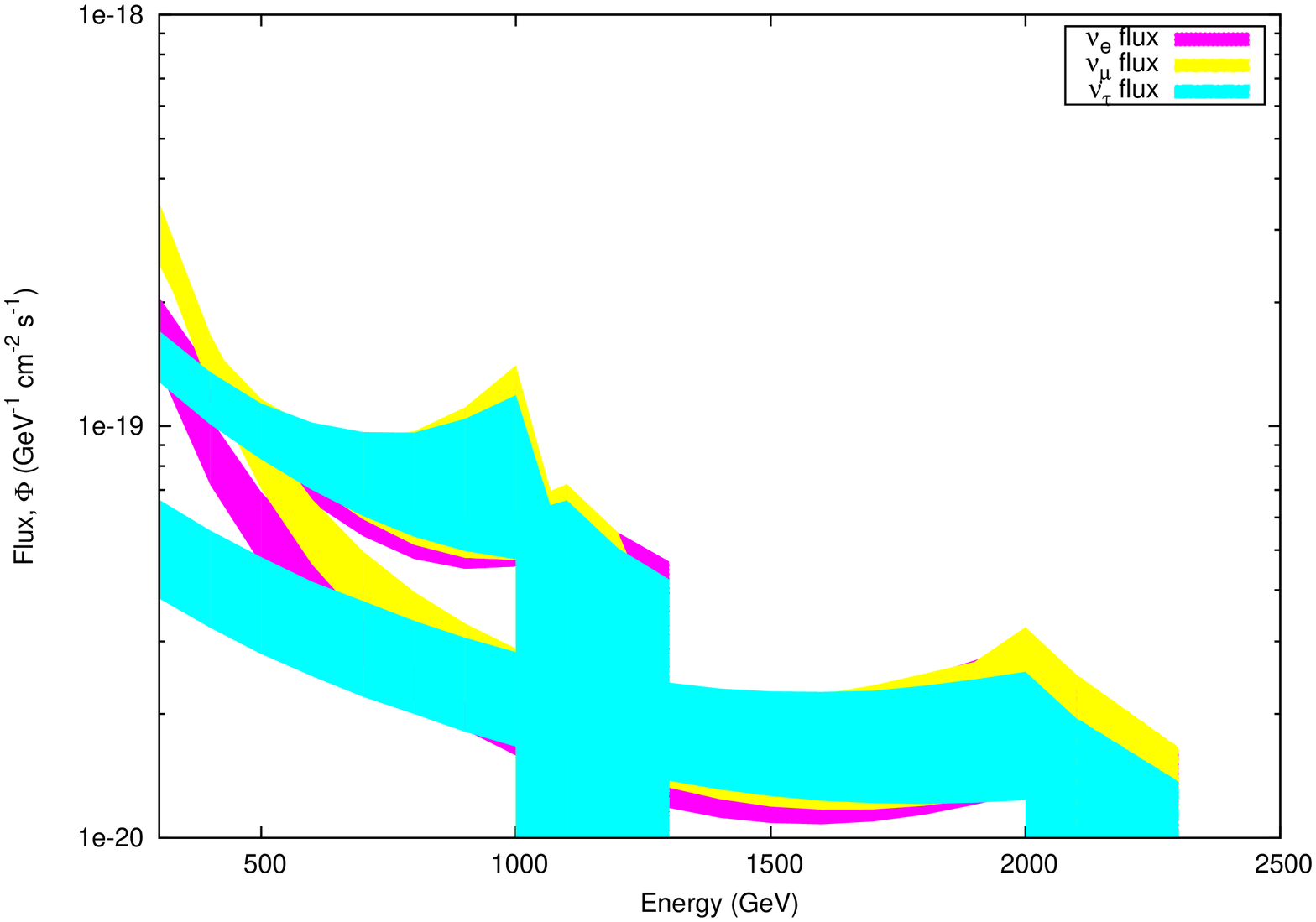}}
\subfigure[\label{fig:nfw3}]{
\includegraphics[width=2.2in,height=2.9in,angle=-90]{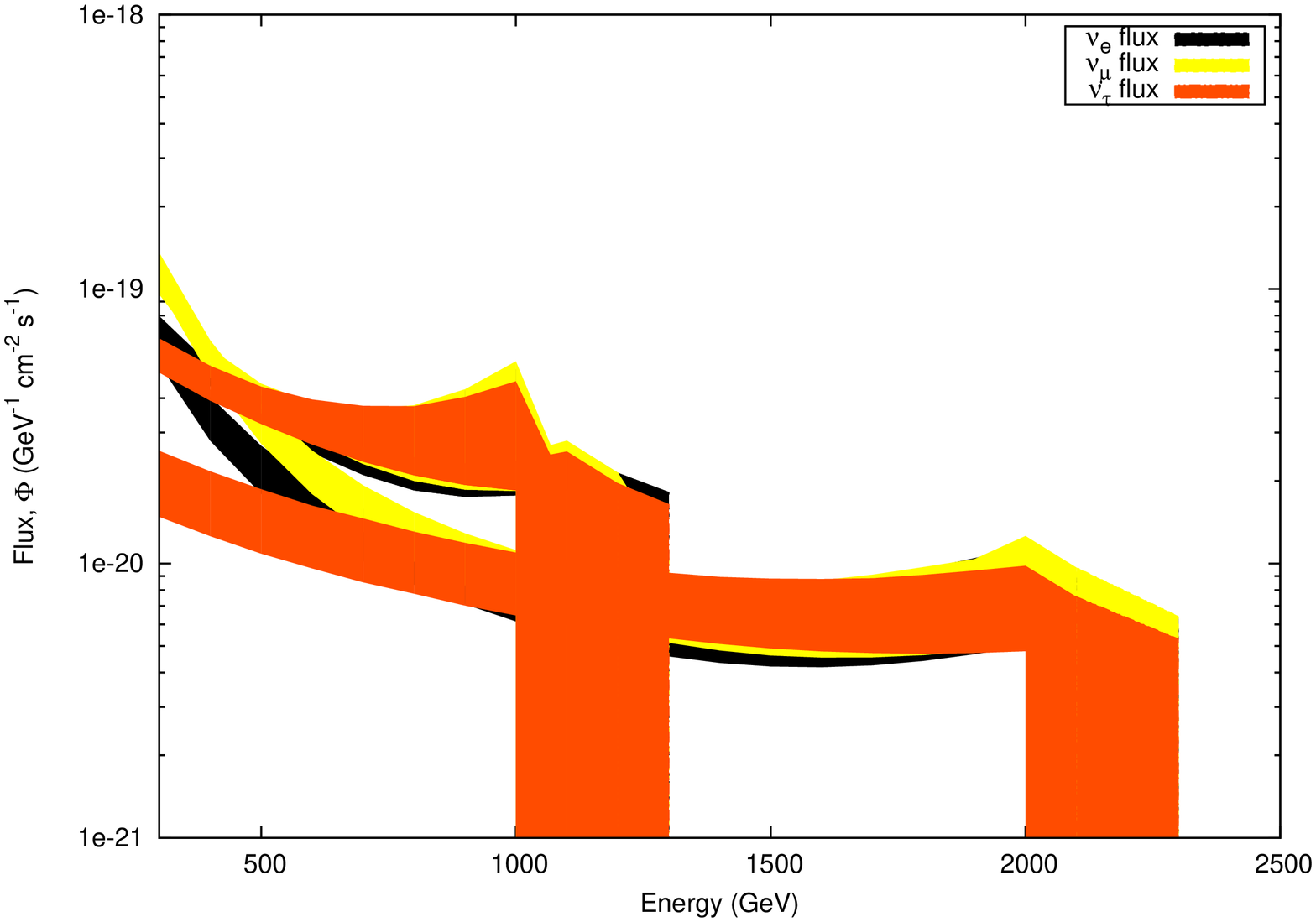}}
\caption{\label{fig:neutrino_nfw} \textit{Neutrino flux of 
three flavours ($\nu_e$, $\nu_\mu$ and $\nu_\tau$) for different
energies
for a) angle of sight, $\theta=0^o$ b) angle of sight, $\theta=30^o$ 
and c) angle of sight, $\theta=60^o$ from the galactic centre respectively.} }
\end{center}
\end{figure}

\begin{figure}[h!]
\begin{center}
\subfigure[NFW]{
\includegraphics[width=2.2in,height=2.9in,angle=-90]{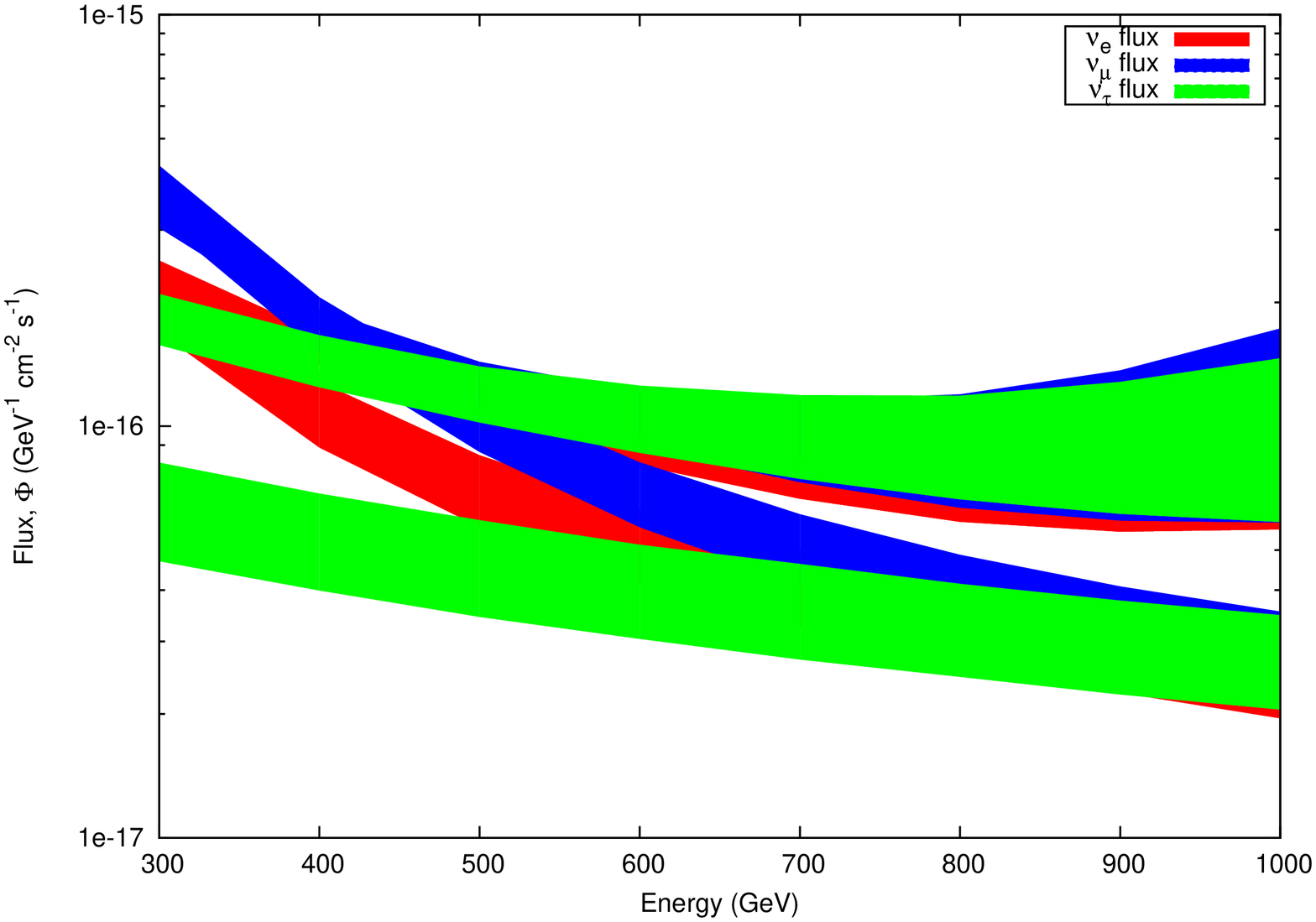}}
\subfigure[Einasto]{
\includegraphics[width=2.2in,height=2.9in,angle=-90]{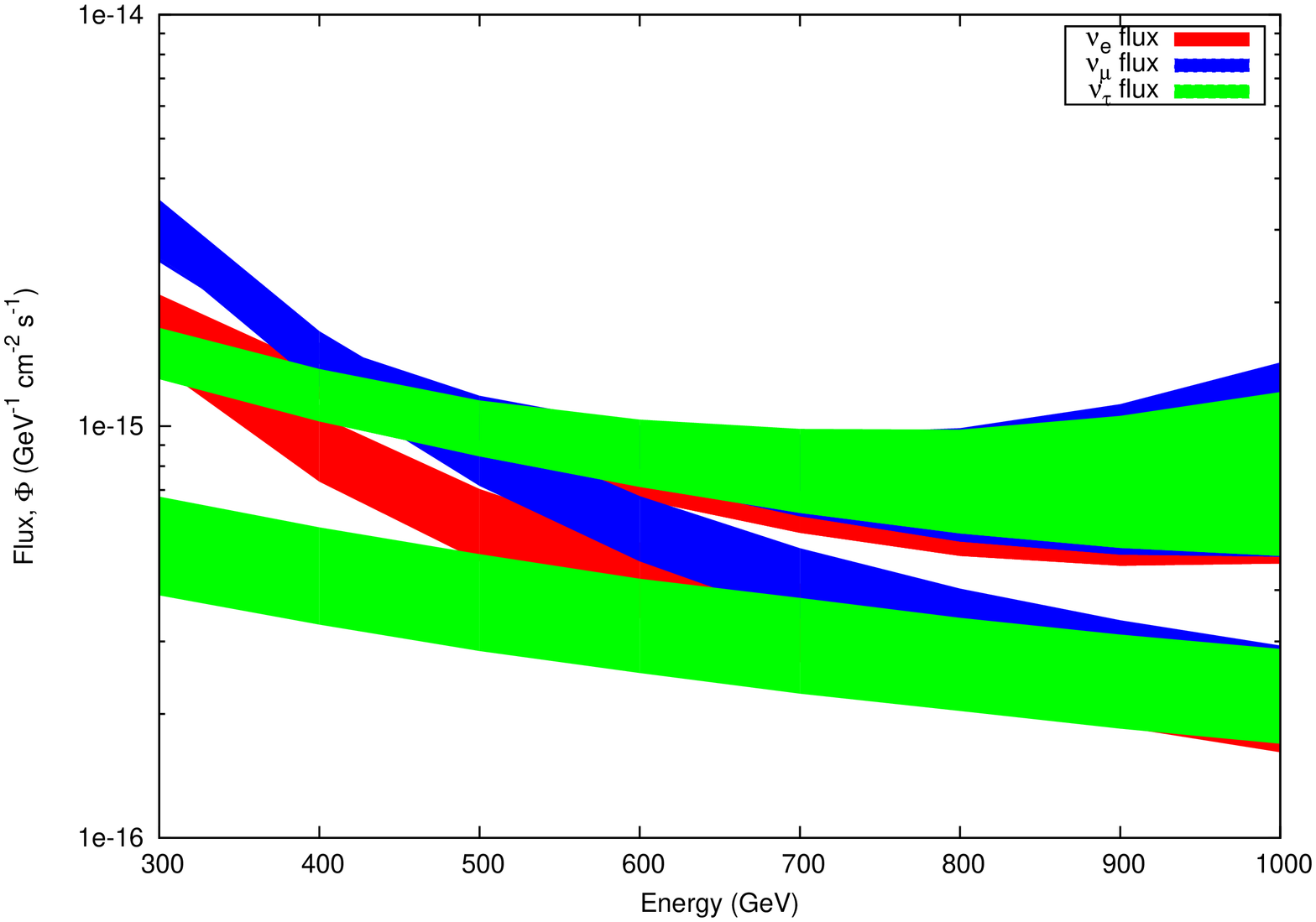}}
\subfigure[Moore]{
\includegraphics[width=2.2in,height=2.9in,angle=-90]{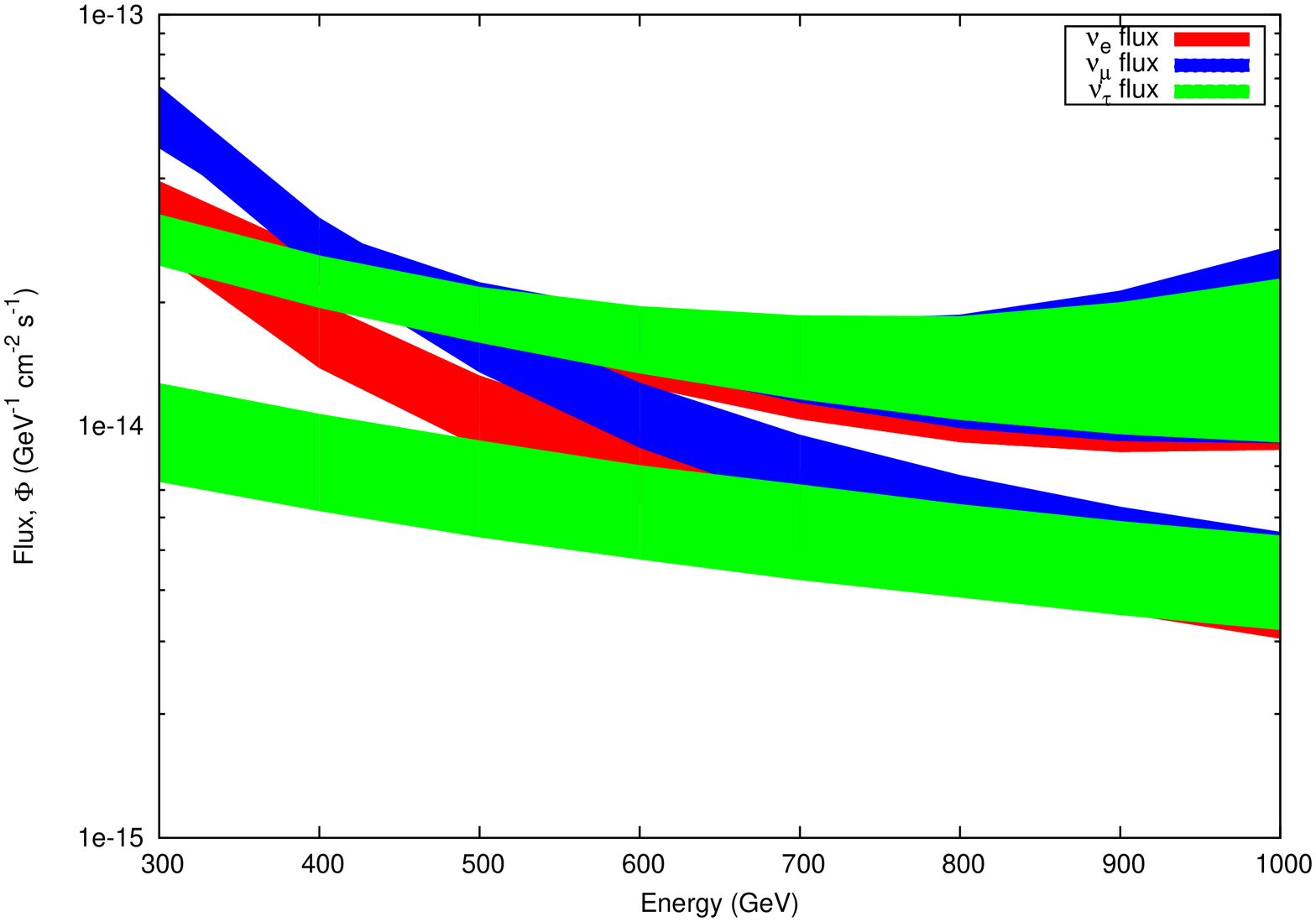}}
\subfigure[Isothermal]{
\includegraphics[width=2.2in,height=2.9in,angle=-90]{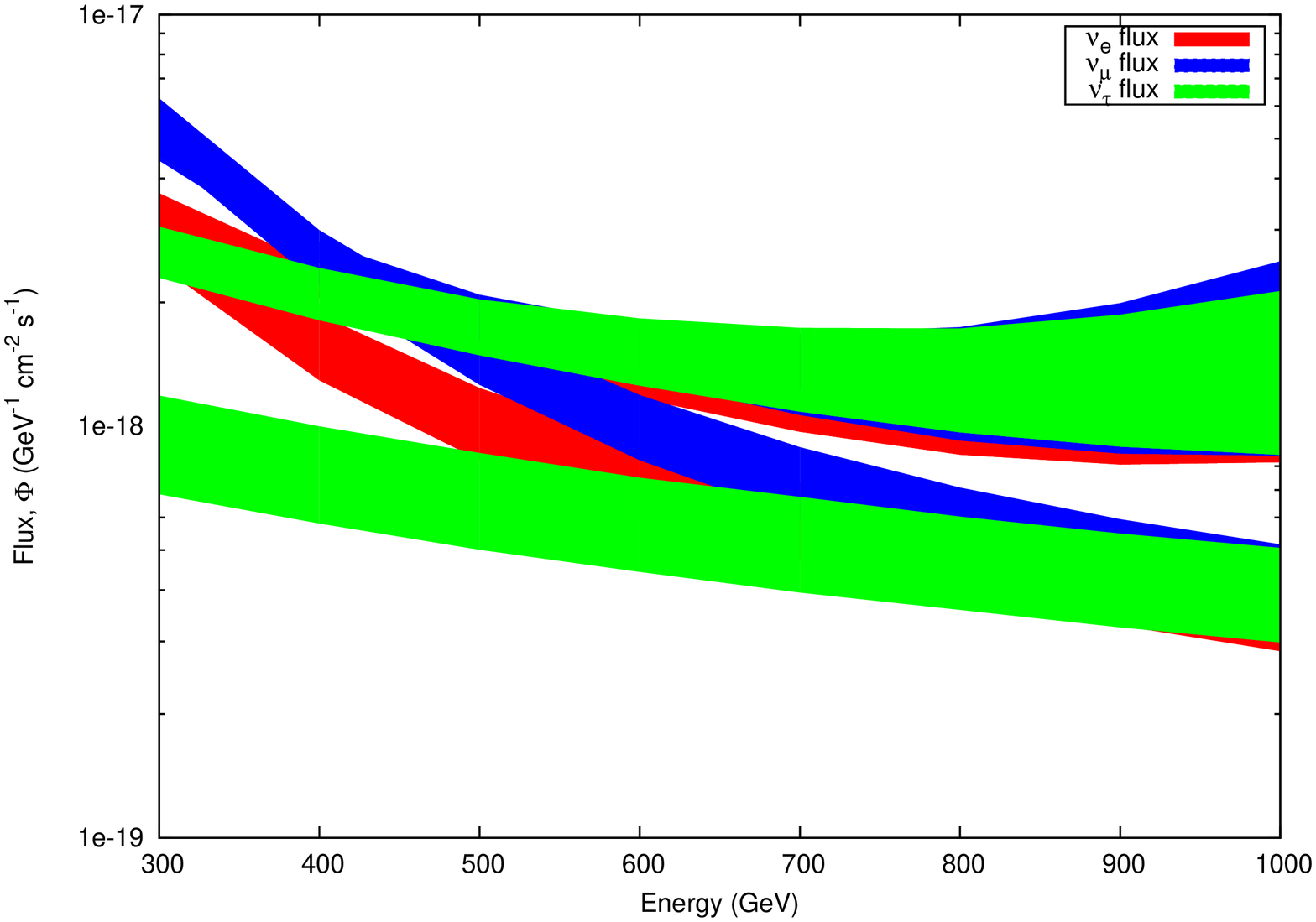}}
\caption{\label{fig:neutrinoflux0} \textit{neutrino flux for 
three flavours ($\nu_e$, $\nu_\mu$ and $\nu_\tau$) for
different energies from the annihilation of dark matter at 
from the galactic centre. The red, blue and green patches
describe the fluxes corresponding to
$\nu_e$, $\nu_\mu$ and $\nu_\tau$
respectively} }
\end{center}
\end{figure}

\begin{figure}[h!]
\begin{center}
\subfigure[NFW]{
\includegraphics[width=2.2in,height=2.9in,angle=-90]{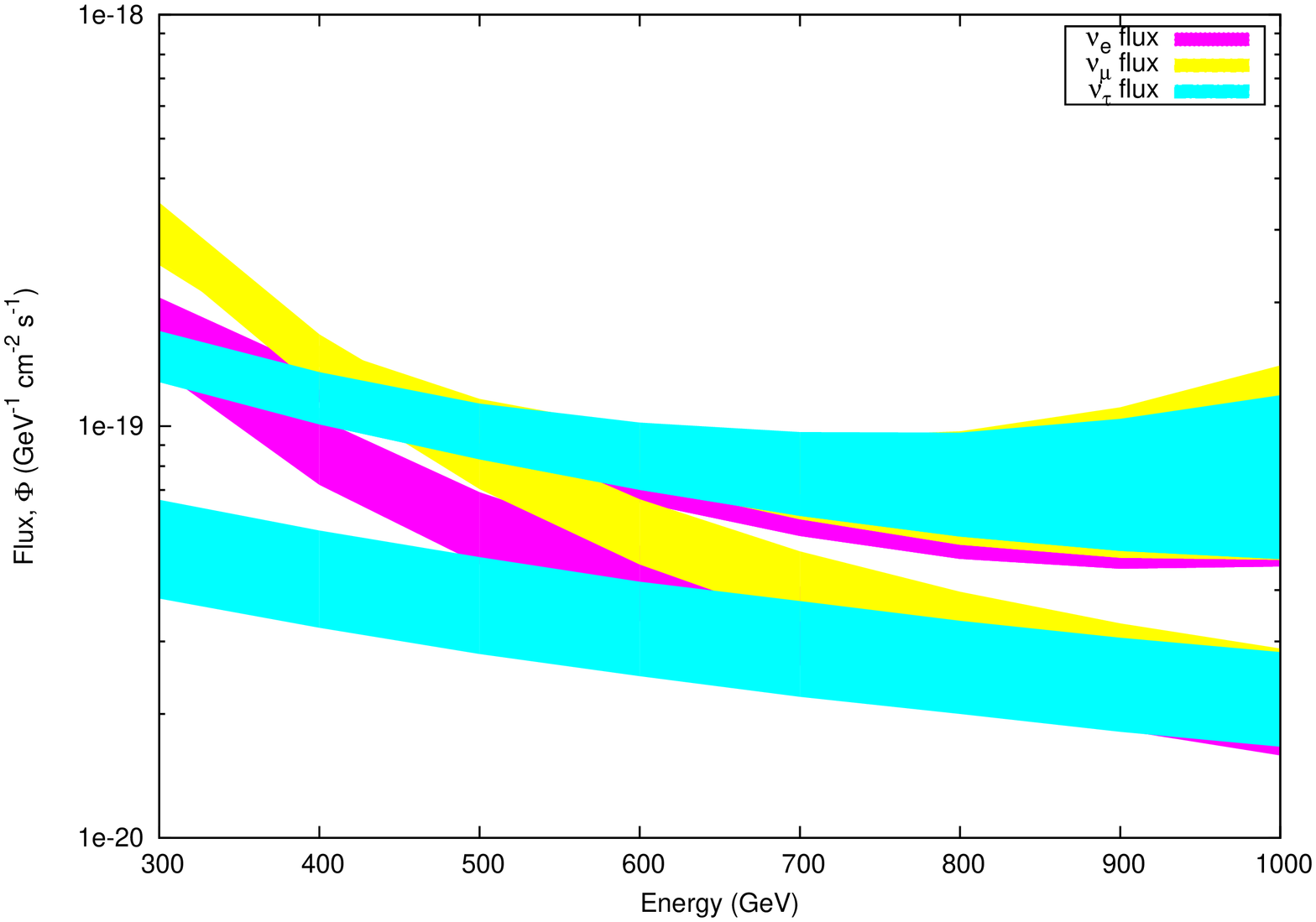}}
\subfigure[Einasto]{
\includegraphics[width=2.2in,height=2.9in,angle=-90]{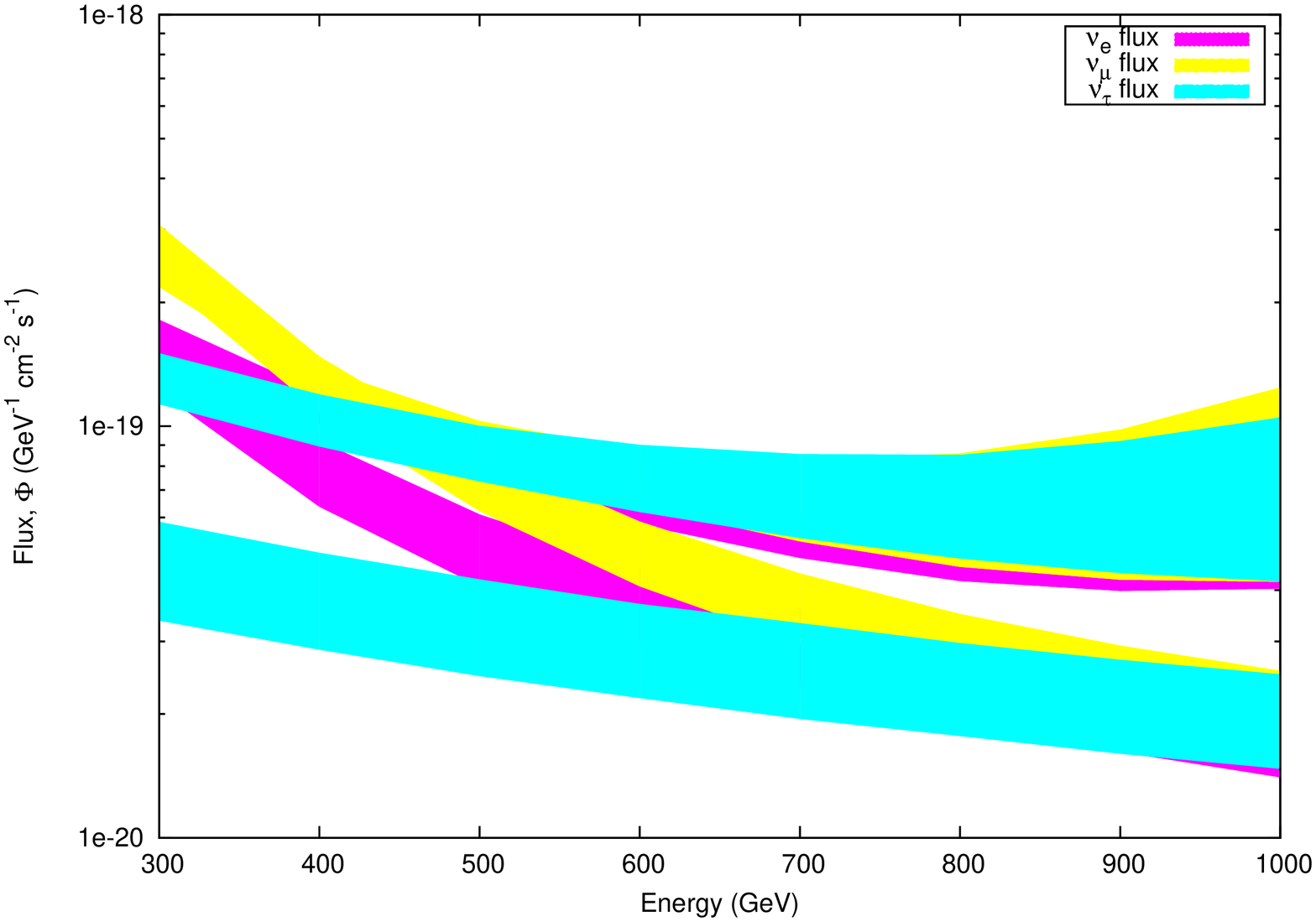}}
\subfigure[Moore]{
\includegraphics[width=2.2in,height=2.9in,angle=-90]{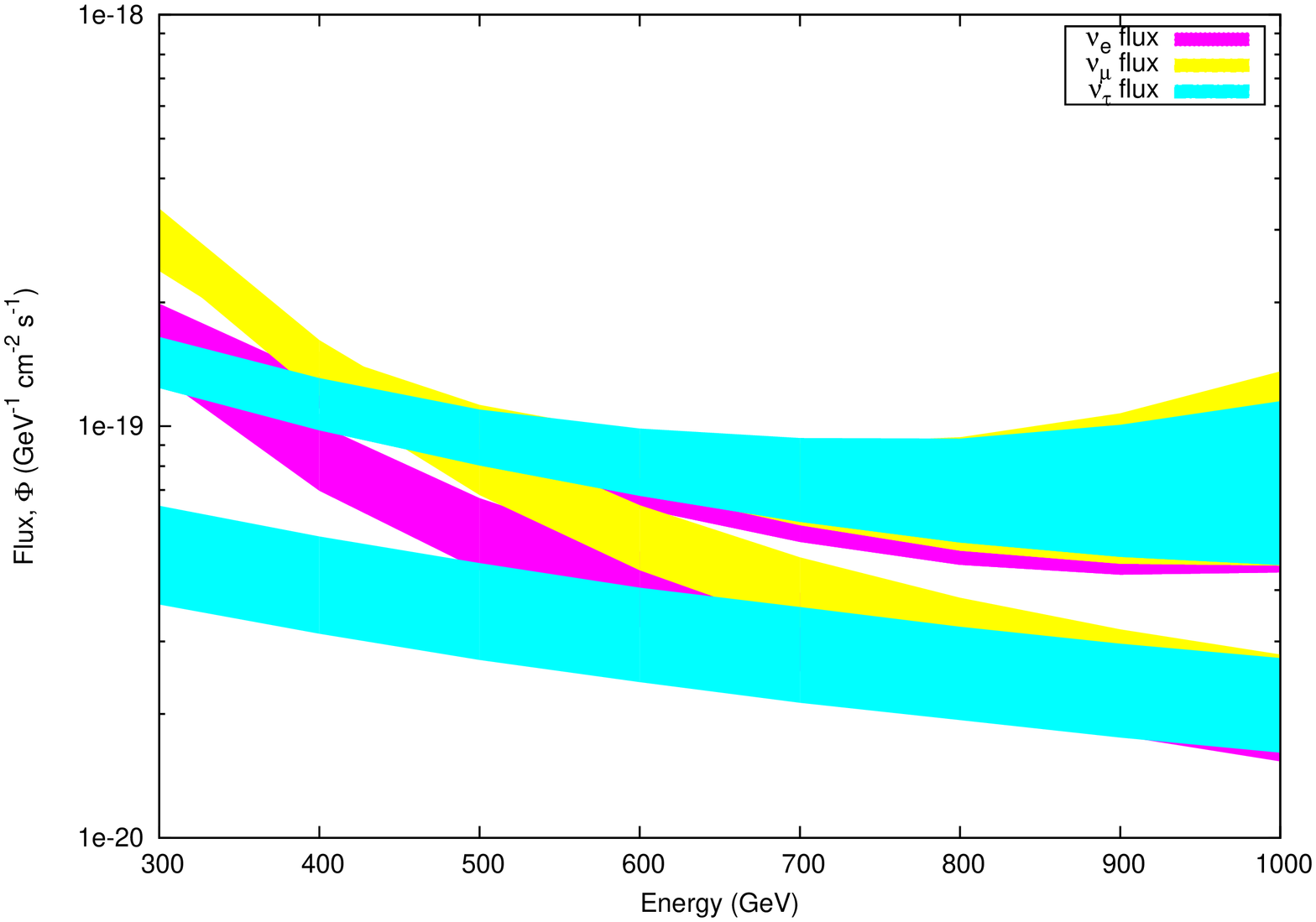}}
\subfigure[Isothermal]{
\includegraphics[width=2.2in,height=2.9in,angle=-90]{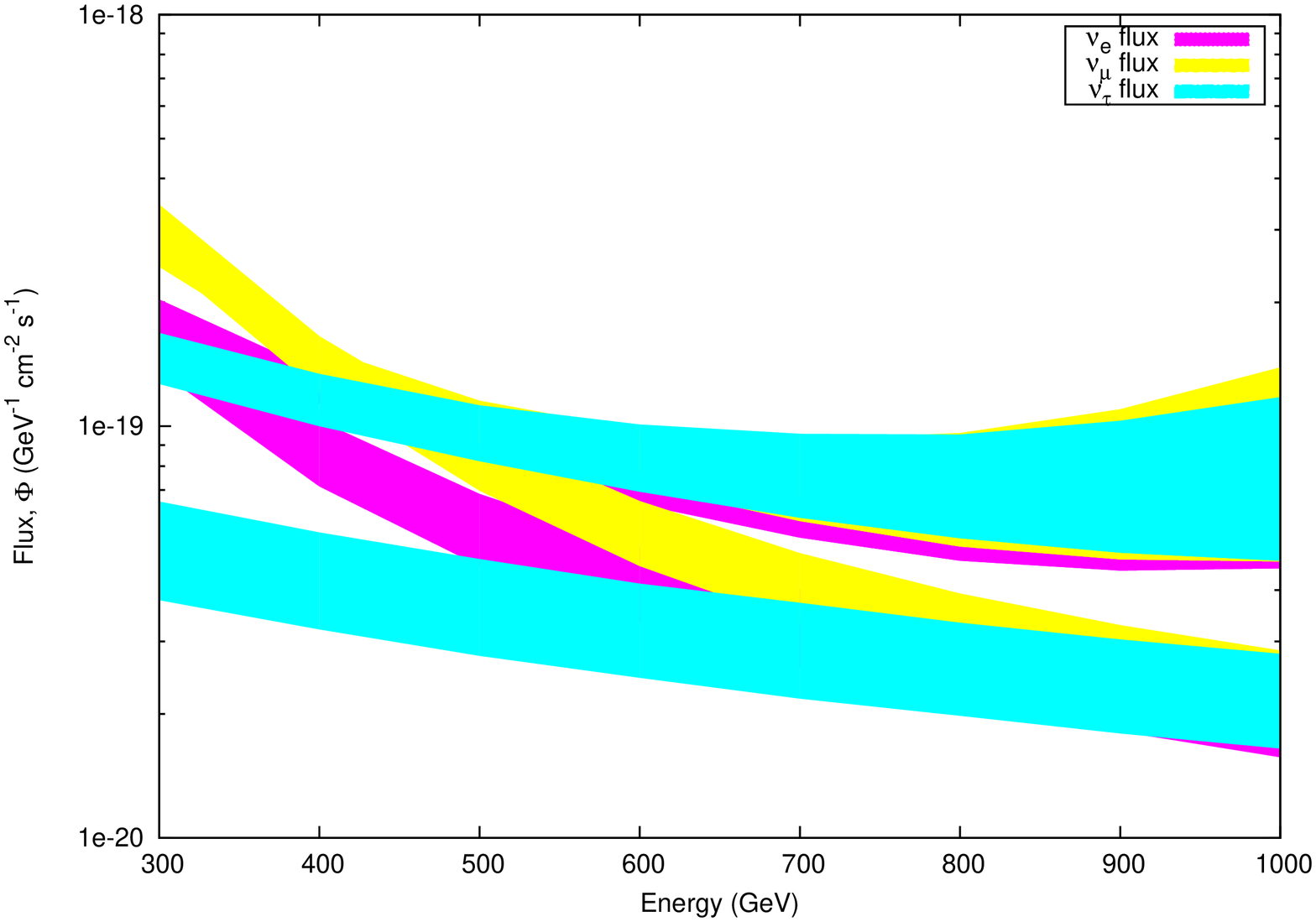}}
\caption{\label{fig:neutrinoflux30} \textit{neutrino flux of 
three flavours ($\nu_e$, $\nu_\mu$ and $\nu_\tau$)
for various energies for angle
of sight, $\theta=30^o$ from the galactic centre. The
pink, yellow and cyan coloured zones
describe the fluxes corresponding to
$\nu_e$, $\nu_\mu$ and $\nu_\tau$
respectively} }
\end{center}
\end{figure}

\begin{figure}[h!]
\begin{center}
\subfigure[NFW]{
\includegraphics[width=2.2in,height=2.9in,angle=-90]{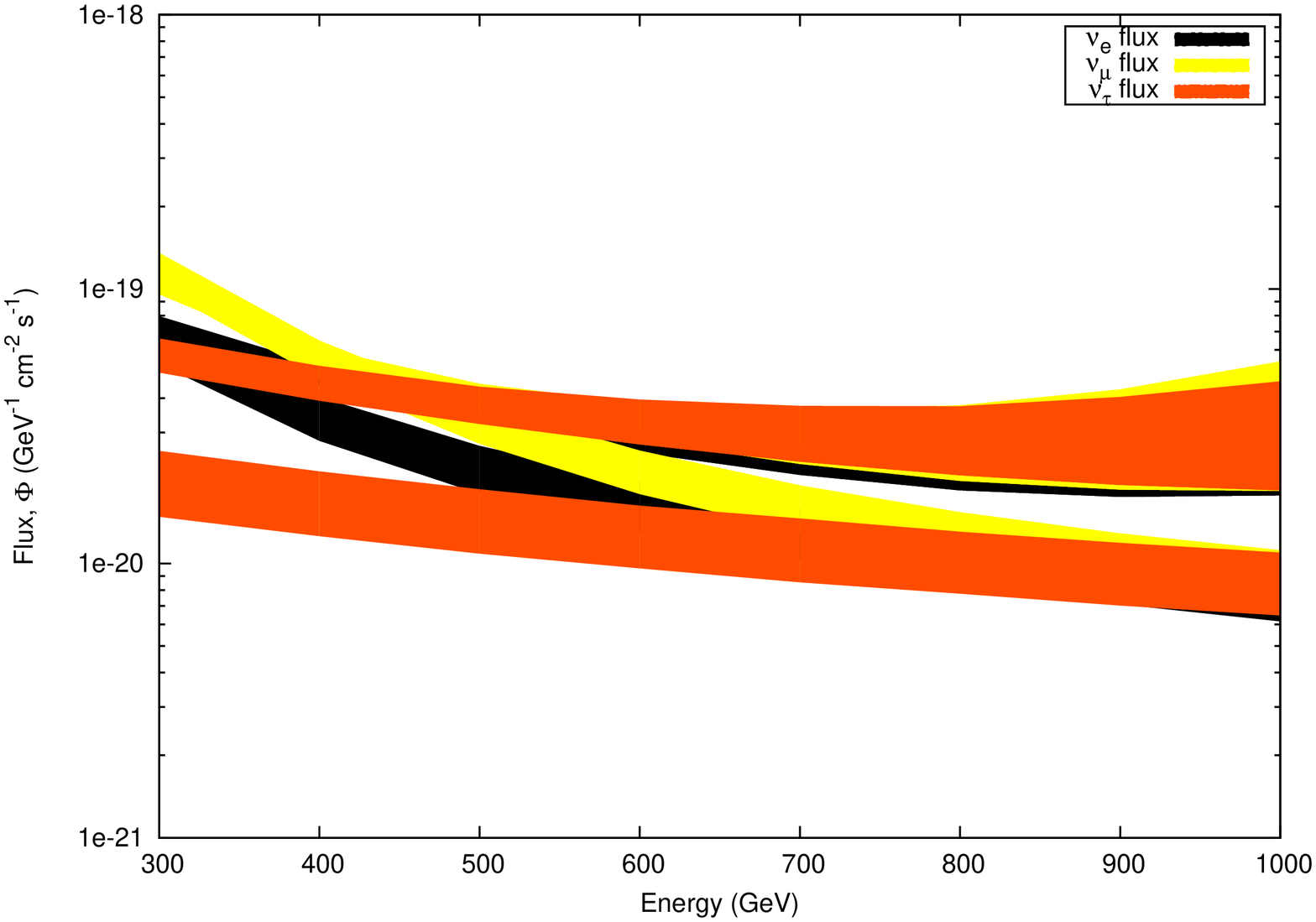}}
\subfigure[Einasto]{
\includegraphics[width=2.2in,height=2.9in,angle=-90]{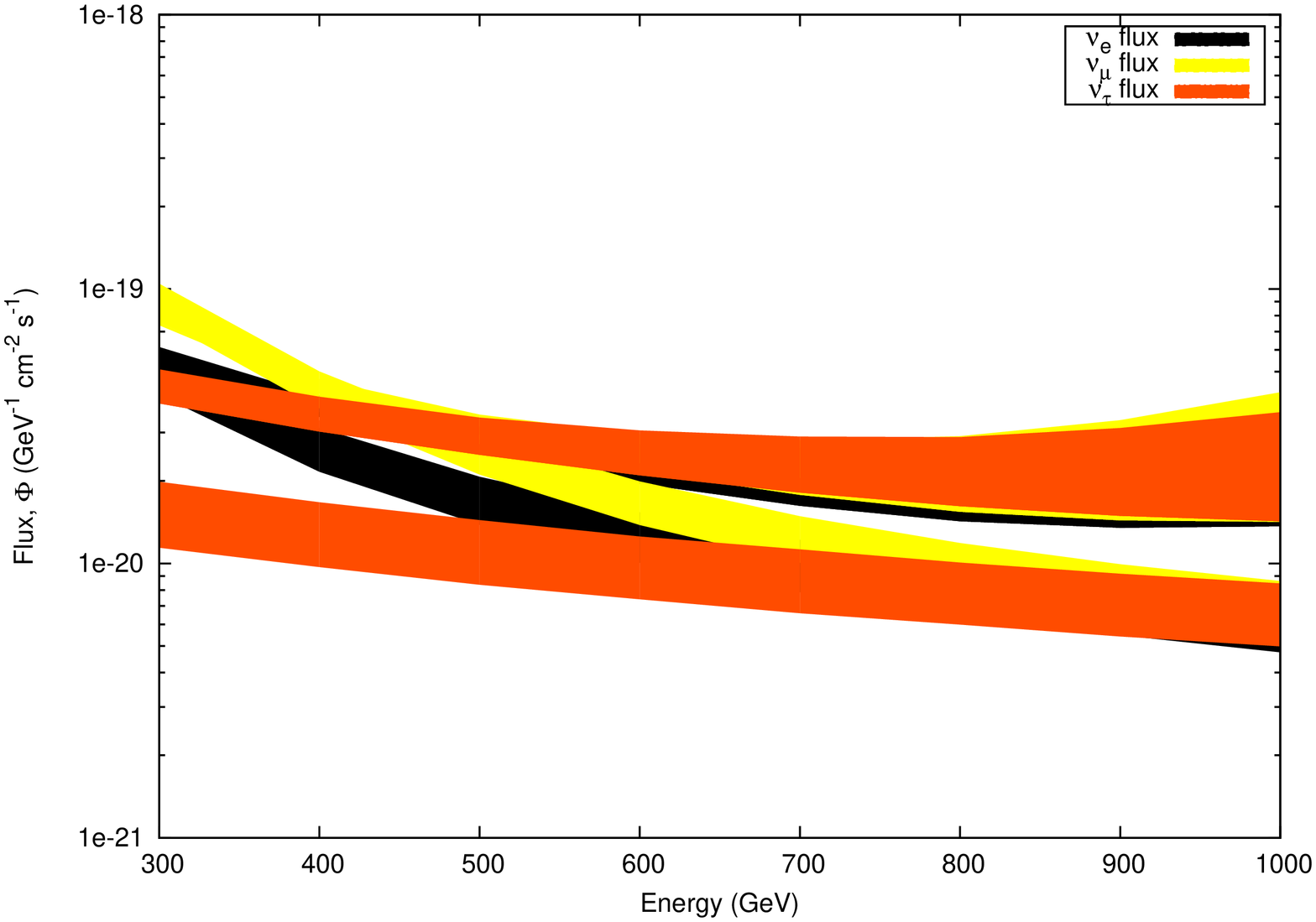}}
\subfigure[Moore]{
\includegraphics[width=2.2in,height=2.9in,angle=-90]{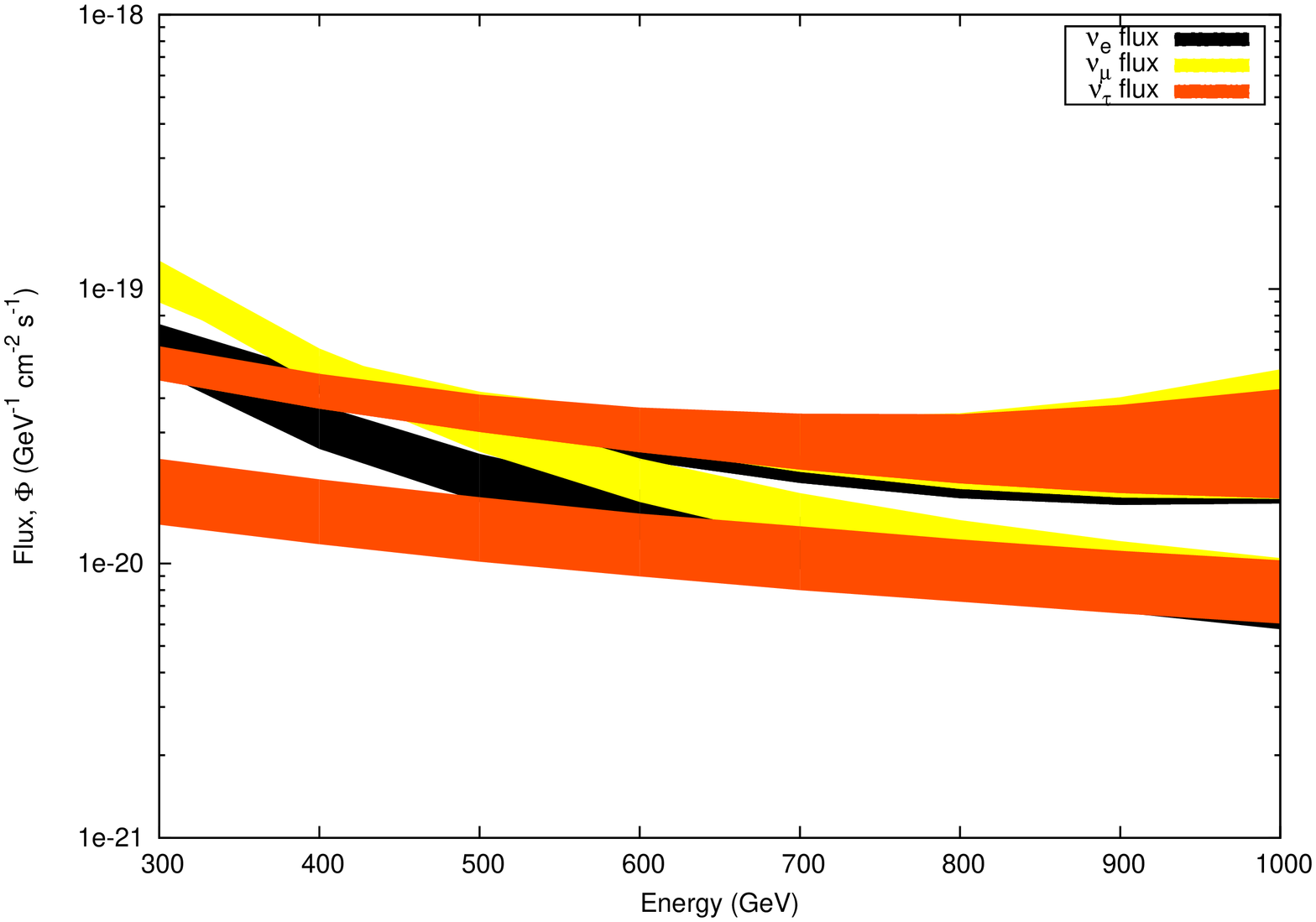}}
\subfigure[Isothermal]{
\includegraphics[width=2.2in,height=2.9in,angle=-90]{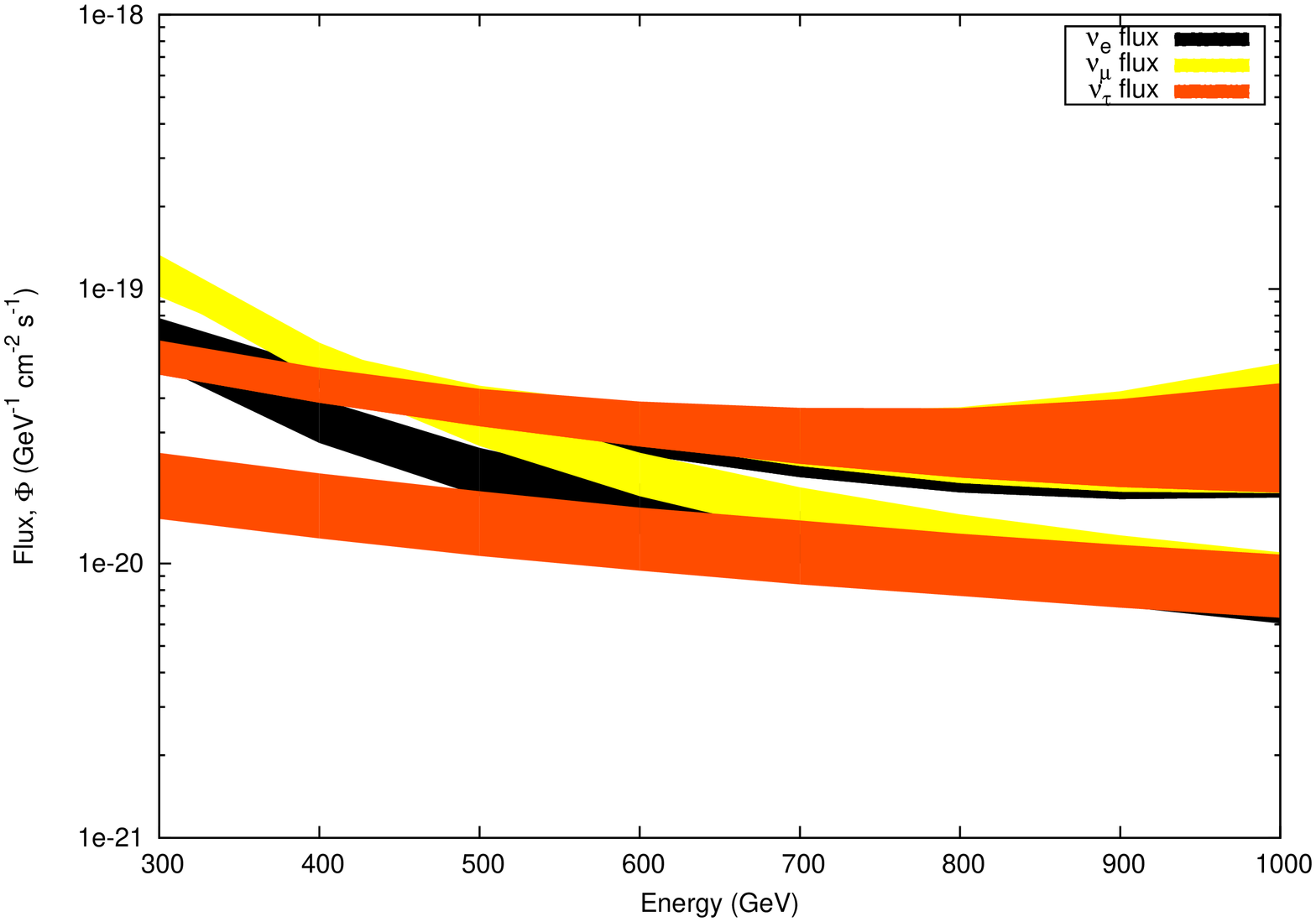}}
\caption{\label{fig:neutrinoflux60} \textit{neutrino flux of 
three flavours ($\nu_e$, $\nu_\mu$ and $\nu_\tau$) for different
energies
for angle of sight, $\theta=60^o$ from the galactic centre.
The black, yellow and orange regions
describe the fluxes corresponding to
$\nu_e$, $\nu_\mu$ and $\nu_\tau$
respectively} }
\end{center}
\end{figure}

We use \texttt{micrOMEGAs} computer code to calculate the neutrino flux 
in the direction of the galactic centre
for all the four halo models considered. The neutrino flux for the 
halo models can be 
obtained using similar equations (Eqs. \ref{flux} - \ref{fhalo}) 
that is used for obtaining $\gamma$-flux. The $\nu$-flux for 
each of the three flavours namely $\nu_e$, $\nu_\mu$ and $\nu_\tau$
are calculated separately for three values of the angle $\theta$ 
(Eq. \ref{flux} and discussions earlier) namely 
$\theta = 0^o,\, \theta = 30^o,\,\theta = 60^o$. The results are furnished
in Figs. \ref{fig:neutrino_nfw} - \ref{fig:neutrinoflux60}.
In Fig. \ref{fig:neutrino_nfw}, we give results only for NFW profile
for the two allowed regions of dark matter mass around 1 TeV 
and around 2 TeV.
We have done similar calculations for other three profiles namely Einasto,
Isothermal and Moore halo profiles. As seen from Fig. \ref{fig:neutrino_nfw}, the
big overlap regions of the plots for the two allowed mass zones reduce
their clarity and readability.
Therefore in Figs. \ref{fig:neutrinoflux0} - \ref{fig:neutrinoflux60}, we plot
the neutrino fluxes for energies upto 1000 GeV for the two allowed dark matter 
mass regions discussed earlier.

The three
figures namely Fig. \ref{fig:neutrinoflux0}, Fig. \ref{fig:neutrinoflux30}
and Fig. \ref{fig:neutrinoflux60} correspond to 
$\theta = 0^o,\, 30^o$ and $60^o$ respectively. In these figures
the $\nu_e$ flux , the $\nu_\mu$ flux and the 
$\nu_\tau$ flux are shown respectively by red, blue and green colours
in Fig. \ref{fig:neutrinoflux0}, 
pink, yellow and turquiose colour labels
in Fig. \ref{fig:neutrinoflux30} and black, yellow
and orange colours in Fig. \ref{fig:neutrinoflux60}
respectively. The two flux regions for each of the neutrino flavours
are for the two different allowed dark matter mass zones in this model 
obtained from WMAP results. 
The $\nu$ flux for different dark 
matter profiles considered here, exhibit similar trends
as for the case of $\gamma$ flux in the sense that the 
flux is more for Moore profile and gradually decreases 
for Einasto, NFW and isothermal profiles. 
 

The neutrinos, while reaching the earth from the galactic centre 
will undergo flavour oscillations, whereby the flux of a particular 
flavour, say $\nu_\mu$, will
be modified on reaching the earth from the galactic centre. Since the 
baseline length $L$ is very large in this case in comparison to oscillation 
length, the osciillation part is averaged out. Thus in the limit 
$L \rightarrow \infty$, the probability
that a neutrino with flavour $\alpha$ will oscillate to flavour 
$\beta$ is given by
\bea
P(\nu_\alpha \rightarrow \nu_\beta;L=\infty) &=& 
\delta_{\alpha\beta} - \sum_{i\neq j} U_{\alpha i}^* U_{\beta i}
U_{\alpha j} U_{\beta j}^* \nonumber \\
&=& |U_{\alpha i}|^2 |U_{\beta i}|^2\,\, ,
\label{prob1}
\eea
where $\alpha$, $\beta$ denote different flavour 
indices, $e$, $\mu$ or $\tau$ and $i,j = 1,2,3$ denote the mass indices of three neutrinos.
In the above, the oscillation part ($\sim \Delta m_{ij}^2 (L/E)$) is 
averaged out due to large $L/E$ ($\sim 10^{13}$ km/GeV).  
The mass-flavour mixing matrix $U$ is denoted by 
\bea
|\nu_\alpha \rangle &=& \sum_i U_{\alpha i} |\nu_i \rangle 
\eea
and 
\bea 
U &\equiv& \left ( \begin{array}{ccc} 
                   U_{e1} &  U_{e2}  & U_{e3} \\
                   U_{\mu 1} &  U_{\mu 2}  & U_{\mu 3} \\
                   U_{\tau 1} &  U_{\tau 2}  & U_{\tau 3} 
                   \end{array} \right )
\eea
In fact $U$ is the usual MNS mixing matrix given by 
\bea
U &=& \left ( \begin{array}{ccc} 
            c_{12}c_{13} & s_{12}s_{13} & s_{13} \\
     -s_{12}c_{23} - c_{12}s_{23}s_{13} & c_{12}c_{23} - s_{12}s_{23}s_{13}
                                        & s_{23}c_{13} \\
     s_{12}s_{23} - c_{12}c_{23}s_{13} & -c_{12}s_{23} - s_{12}c_{23}s_{13}
                                        & c_{23}c_{13} 
        \end{array} \right )\,\, .
\label{mns}
\eea  
In the above, $s$, $c$ denote $\sin\theta$, $\cos\theta$ respectively 
and $\theta_{12}$, $\theta_{23}$ and $\theta_{13}$ are three mixing 
angles for three neutrino species. We consider here no CP violation 
in neutrino sector. 
From Eq. \ref{prob1} probability $P$ can be written as
\bea
P \equiv XX^T
\label{prob2}
\eea
where the matrix $X$ is given by 
\bea
X &\equiv&  \left ( \begin{array}{ccc}
                   |U_{e1}|^2 &  |U_{e2}|^2  & |U_{e3}|^2 \\
                   |U_{\mu 1}|^2 &  |U_{\mu 2}|^2  & |U_{\mu 3}|^2 \\
                   |U_{\tau 1}|^2 &  |U_{\tau 2}|^2  & |U_{\tau 3}|^2
                   \end{array} \right )\,\, .
\label{prob3}
\eea 
Hence, the oscillated flux of the neutrinos (of three flavours) 
at the detector is given by 
\bea 
\left ( \begin{array}{c} \phi_{\nu_e} \\ \phi_{\nu_\mu} \\ \phi_{\nu_\tau}
\end{array} \right ) &=& XX^T 
\left ( \begin{array}{c} \phi_{\nu_e}^0 \\ \phi_{\nu_\mu}^0 \\ 
\phi_{\nu_\tau}^0
\end{array} \right )\,\, ,
\label{prob4}
\eea
where the quantities in the RHS with superfix 0 denote the initial 
neutrino fluxes.  

In the present work we estimate the muon yield for such a $\nu_\mu$
flux from galactic centre at ANTARES neutrino detector. ANTARES is a deep
sea neutrino telescope and is basically a water Cerenkov detector, 
which detect the neutrinos by detecting the Cerenkov light 
of a charged lepton that is produced by the charged current scattering of 
neutrino off the sea water. The telescope consists of several vertical strings
of around 350 metres long, each of which is fixed with 75 optical
modules containing photomultiplier  
tubes. The strings are installed at the Mediterranian sea bed 
at a depth of around 2.5 Km off the French coast of Toulon.   
Designed to detect neutrinos with high energy ($\sim 100$ GeV to $\sim 100$
TeV) of generally cosmic origin, this telescope looks in the direction of 
southern hemisphere. In fact due its position, ANTARES is very much suitable
for observing the galactic plane and the galactic centre. 
In the present context, the dark matter mass 
from AMSB model that is allowed by WMAP, is in the region of
$\sim 1$ TeV - $\sim 2$ TeV and since neutrinos from the annihilation 
of such dark matter at galactic centre is being studied, 
this telescope is best suited for the purpose. 
   
The spectrum of muon yield, $\Phi_\mu (E_{\nu_\mu})$, 
for different enrgies $E_{\nu_\mu}$ at ANTARES can be estimated
using the relation
\bea
\Phi_\mu (E_{\nu_\mu}) &=& \phi_{\nu_\mu} A_{{\rm eff},\nu} (E_{\nu_\mu})\,\, ,
\eea 
where, $A_{{\rm eff},\nu}$ is the neutrino effective area for 
ANTARES telescope and is obtained from 
Ref. \cite{antares}. 

\begin{figure}[h]
\begin{center}
\includegraphics[width=2.7in,height=4.5in,angle=-90]{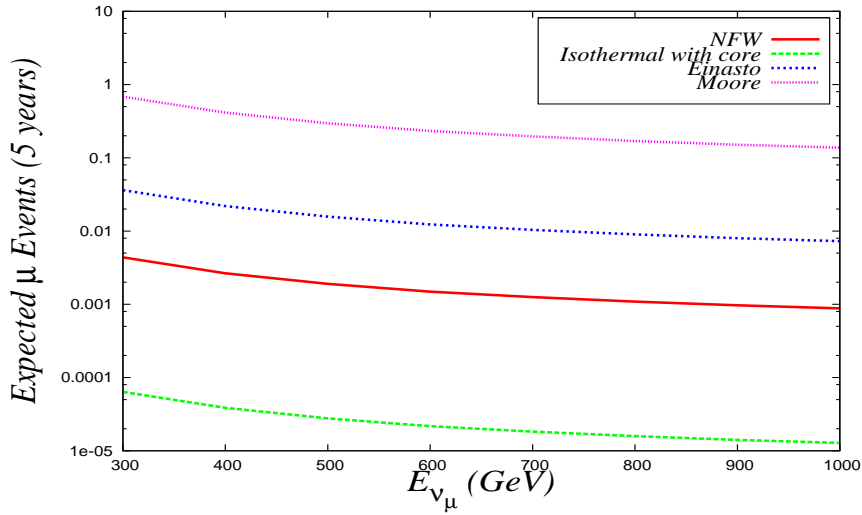}
\caption{\label{antares} \textit{Estimated $\mu$ events for five
year run at ANTARES neutrino telescope for different 
$\nu_\mu$ energies obtained from dark matter annihilations 
at the galactic centre in the framework of mAMSB model.}}
\end{center}
\end{figure}

The $\nu_\mu$ flux, $\phi_{\nu_\mu}$, 
at the earth from the galactic
centre is calculated using Eqs. \ref{prob1} - \ref{prob4} 
with $\phi_{\nu_\mu}^0$, the flux at the source,  
given in Fig. \ref{fig:neutrinoflux0}
for different halo profiles. Note that, $\phi_{\nu_\mu}^0$ 
at galactic centre is considered only for the case $\theta =0$ (see
earlier in this section) in the present calculation.  
The values of three neutrino mixing angles in Eq. \ref{mns} are taken to 
be $\theta_{12} = 34.0^o$ \cite{theta12}, $\theta_{23} = 46.1^o$ 
\cite{theta12} and $\theta_{13} = 9.2^o$ \cite{theta13}. 
The results for estimated yield of muon spectrum $\Phi_\mu (E_{\nu_\mu})$ 
at ANTARES is shown in Fig. \ref{antares} for all the four halo 
profiles considered. The estimates are shown for 5 year run of the telescope.
It is seen from Fig. \ref{antares}, that while NFW profile predicts very large
yield, the same using the isothermal profile is rather low. The NFW profile has 
a cuspy structure whereas the isothermal profile gives a flat halo.

If ANTARES detects $\nu_\mu$ from galactic centre then the $\mu$ signal
from such detection can be compared with the results given in Fig. \ref{antares}
for different DM halo profiles. Such comparison could readily give an idea
of the dark matter halo profile as also the viability of the present model for
DM candidate. Thus these observations can be used to probe the nature of the
halo profile and the particle physics model of the dark matter as well.

\section{Summary and Conclusion}
In this work, we have investigated the phenomenological implications of 
dark matter coming from a very
well known SUSY breaking model, namely
minimal anomaly mediated supersymmetry breaking 
(mAMSB) model. The suitable candidate in this model is 
the neutralino stabilised by the conservation of R-parity in SUSY theory. 
We have
randomly scanned the parameter space of this model within the
theoretical bounds of the parameter 
space of this model 
and for each point in paremeter space, we obtain a
neutral stable candidate (neutralino) of dark matter. 
In doing so, latest bound on the chargino mass as given by the
ATLAS collaboration is adopted. 

The mass of the LSP neutralino in the present scenario is obtained
in two regions of which one is around 1 TeV and
the other is at a somewhat higher range of $\sim$ 2 TeV.
We have checked that these neutralinos are predominantly of wino type.
The measure of the naturalness which is expressed in terms of the commonly used
fine tuning parameters  
are obtained for the constrained neuralino masses (vide earlier) in the present scenario,
$$
\frac{\delta {M_Z}^2}{{M_Z}^2} ({\mu}^2) \sim 10^4,\,\,\,\,\,\,\, 
\frac{\delta {M_Z}^2}{{M_Z}^2} (B_{\mu}) \sim 10^3
$$
$$
\frac{\delta M_t}{M_t} ({\mu}^2)  \sim 10^4, \,\,\,\,\,\, 
\frac{\delta M_t}{M_t} ({\mu}^2) \sim 10^3 \,\, ,
$$

where the symbols have their usual significance.

We calculate the relic densities of such neutralinos 
and compare them with WMAP bounds to obtain the mass zones of these
mAMSB neutralinos that satisfy WMAP limits.
The allowed parameters determined from such constraints are then used 
to study the direct and indirect detections of the proposed dark matter
candidate, neutralino, in the present mAMSB model. 

The scattering cross sections for the 
dark matter particles scattered off the nucleus of the detecting 
material are determined by the nuclear form factors and the 
dark matter-nucleon coupling. The two types of cross sections, namely 
spin independent (zero nuclear spin at the ground state) 
and spin dependent, are determined by the different form factors and 
dark matter-nucleon couplings. 
We calculate both the spin dependent and spin independent
scattering cross sections with the constrained zone(s) of the present 
neutralino dark matter parameter space 
and hence compared our results with several recent ongoing direct
detection experimental results. The calculated 
mass-cross sections, thus calculated, are found to be below  
the upper limits of many well
known experiments. 
From experimental point of view, 
the future advanced direct detection
techniques may probe those regions in mass-cross section plane 
given from the model.

We have computed the gamma ray flux coming from the galactic centre and its neighbourhood,
considering that they are produced from the annihilation of 
dark matter in mAMSB model. For this reason, 
we have taken several
well known theoretically motivated dark matter halo profiles such 
as NFW profile, 
Moore profile, isothermal profiles with core and Einasto
profiles and calculate the flux coming
from different positions of the halo plane. As the allowed mass of the 
neutralino (dark matter) is high ($\sim$ few GeV to $\sim$ $10^3$ GeV), 
the energies of
the gamma rays from dark matter annihilations are also of that 
order. Therefore high energy gamma ray search
experiments may verify the present model. 
For this purpose, we have chosen the HESS experiment and compared our 
results for different halo profiles considered, with the 
observed $\gamma$-flux of this experiment.
In the passing we also mention that another water $\hat{C}$erenkov detector
namely HAWC (High-Altitude Water Cherenkov Gamma-Ray Observatory) \cite{hawc} near
Puebla, Mexico can also detect gamma ray annihilation signal in the energy domain
of 1 $\sim$ 2 TeV. But as mentioned, in this work we consider only HESS experiment.
We find that the $\gamma$-fluxes for
non-cuspy profiles like isothermal profile (flat) and Einasto profile,
are orders below the HESS results whereas the cuspy profiles like 
Moore profile overestimate the HESS result. Calculations using 
the other cuspy profile, namely the NFW profile requires a boost
of $\sim 10^3$ for comparison with HESS results. The Moore profile
has an asymptotic slope, $\alpha = 1.5$, 
while the same for the NFW profile is $\alpha = 1.0$. 
Thus the former is steeper than the latter.
Cuspy nature appears to influence
the result. It is still a matter 
of investigation to understand whether
halo profile at the galactic centre has a 
flat profile or a steep profile. The present analysis, 
within the framework of mAMSB model for 
dark matter candidate, seems to suggest
that the cuspy nature of the profile appears to explain the HESS data
better than the flat ones. We also like to add that we performed 
similar calculations with another flat halo profile namely Burkert 
profile \cite{burkert1,burkert2}, but the calulated $\gamma$ 
flux is found to be even below than what is obtained for isothermal profile.         

Different flavours of neutrinos from the dark matter annihilation 
at galactic centre are also addressed in the present work. 
The flux and detection of muon species of such neutrinos
are calculated for the neutralino dark matter in mAMSB model.
Given the masses of such dark matter candidates the energies
of such neutrinos will also be in the range GeV to TeV.  
The location of the galactic centre with respect to earth
is downwards. The high energetic muon neutrinos may  
produce muons by the charged current scattering off ice or water 
and may be detected by their Cerenkov lights. 
We calculate the fluxes of neutrinos of different flavours
due to annihilations of dark matter
when viewed in the direction of the galactic centre as also at the other two chosen 
positions in its neighbourhood. The results are shown for the four halo 
profiles considered. In order to estimate the detection yield
of such neutrinos in a terrestrial neutrino observatory, we have 
chosen the ANTARES under sea detector and calculated the muon yield
for muon neutrinos from galactic centre for all the four 
halo profiles considered. The calculations of neutrinos in case of 
different halo profiles also exhibit
similar trend as those for the calculation of $\gamma$ flux.   

The value of thermal average of the squared halo density,
$\langle\rho^2(r)\rangle$ is generally greater     
than $(\langle\rho(r)\rangle)^2$ due to the influence of     
a probable clumpy structure of dark matter
halo profile, $F_c(\rm r)$, which is related to
dark matter halo profile by,
\begin{equation}
 \langle\rho^2({\tilde r})\rangle = \rho^2_0F^2_{halo}(r)F_c(r)
\end{equation}
The clump structure of dark matter halo gives rise to enhancement factor. 
In the present study of different models of galactic halo structures,
we did not consider any clumpy halo of       
dark matter. This study is for posterity.     

The WMAP allowed zone(s) for the mAMSB model for dark matter, 
are around ($\sim$ 1 TeV and $\sim$ 2 TeV)
which are high in mass regime like Kaluza-Klein dark matter.
The future collider experiment may verify their existence.

\section{Acknowledgements}

The authors thank Pijushpani Bhattacharjee and Pratik Majumdar for some useful discussions.

\end{document}